%% file: mackenty.tex
\documentclass[preprint]{aastex}
\newcommand{\ha}{\mbox{H$\alpha$}}
\newcommand{\hb}{\mbox{H$\beta$}}
\newcommand{\hei}{\mbox{\ion{He}{1}~$\lambda$6678}}
\newcommand{\oiiir}{\mbox{[\ion{O}{3}]~$\lambda$5007}}
\newcommand{\oiiil}{\mbox{[\ion{O}{3}]~$\lambda$4959}}
\newcommand{\oiii}{\mbox{[\ion{O}{3}]~$\lambda$4959+5007}}
\newcommand{\niir}{\mbox{[\ion{N}{2}]~$\lambda$6584}}
\newcommand{\nii}{\mbox{[\ion{N}{2}]~$\lambda$6548+6584}}
\newcommand{\sii}{\mbox{[\ion{S}{2}]~$\lambda$6717+6731}}
\newcommand{\siil}{\mbox{[\ion{S}{2}]~$\lambda$6717}}
\newcommand{\siir}{\mbox{[\ion{S}{2}]~$\lambda$6731}}
\newcommand{\wha}{\mbox{$W$(H$\alpha$)}}
\newcommand{\whb}{\mbox{$W$(H$\beta$)}}
\newcommand{\woiiir}{\mbox{$W$([\ion{O}{3}]~$\lambda$5007)}}
\newcommand{\wwr}{\mbox{$W$(WR~$\lambda$4686)}}
\newcommand{\lwr}{\mbox{$L$(WR~$\lambda$4686)}}
\newcommand{\lhb}{\mbox{$L$(H$\beta$)}}
\singlespace
\slugcomment{To appear in the Astronomical Journal}
\lefthead{MacKenty et al. 2000}
\righthead{HST/WFPC2 and VLA Observations of the Ionized Gas in NGC 4214}
\begin{document}
\newcommand{\etal}{{\it et al.}\ } 

\title{HST/WFPC2 and VLA Observations of the Ionized Gas in the Dwarf Starburst
Galaxy \objectname{NGC 4214}\altaffilmark{1}} 
\shorttitle{The Ionized Gas in \objectname{NGC 4214}}

\author{John W. MacKenty\altaffilmark{2},
        Jes\'us Ma\'{\i}z-Apell\'aniz\altaffilmark{2,3},
        Christopher E. Pickens\altaffilmark{4},
        Colin A. Norman\altaffilmark{2,4}, \and
        Nolan R. Walborn\altaffilmark{2}}


\altaffiltext{1}{Based on observations with the NASA/ESA {\em Hubble Space 
Telescope} and the NRAO {\em Very Large Array}. The HST observations were
obtained at the Space Telescope Science Institute, which is operated by
the Association of Universities for Research in Astronomy, Inc. under NASA 
contract No. NAS5-26555. The National Radio Astronomy Observatory is a facility 
of the National Science Foundation operated under cooperative agreement by 
Associated Universities, Inc.}

\altaffiltext{2}{Space Telescope Science Institute, 3700 San Martin Drive, 
Baltimore, MD 21218, U.S.A.}

\altaffiltext{3}{Laboratorio de Astrof\'{\i}sica Espacial y F\'{\i}sica
Fundamental-INTA, Apdo. Postal 50727, E-28080 Madrid, Spain.}

\altaffiltext{4}{Henry A. Rowland Department of Physics and Astronomy, The 
Johns Hopkins University, Baltimore, Maryland 21218, U.S.A.}



\begin{abstract}

 	We present new \ha\ and \oiiir\ narrow band images of the starbursting 
dwarf galaxy \objectname{NGC 4214}, obtained with the Wide Field and Planetary 
Camera (WFPC2) onboard the \textit{Hubble Space Telescope} (HST), together with 
VLA observations of the same galaxy. The HST images resolve features down to 
physical scales of $2-5$ pc, revealing several young ($< 10$ Myr) star forming 
complexes of various ionized gas morphologies (compact knots, complete or 
fragmentary shells) and sizes ($\sim 10-200$ pc). Our results are consistent 
with a uniform set of evolutionary trends: The youngest, smaller, filled 
regions that presumably are those just emerging from dense star forming clouds, 
tend to be of high excitation and are highly obscured. Evolved, larger 
shell-like regions have lower excitation and are less extincted due of the 
action of stellar winds and supernovae. In at least one case we find evidence 
for induced star formation which has led to a two-stage starburst. Age estimates
based on \wha\ measurements do not agree with those inferred from 
wind-driven shell models of expanding \ion{H}{2} regions. The most likely 
explanation for this effect is the existence of an $\approx$ 2 Myr delay in the 
formation of superbubbles caused by the pressure exerted by the high density 
medium in which massive stars are born. We report the detection of a supernova 
remnant embedded in one of the two large \ion{H}{2} complexes of NGC 4214.
The dust in NGC 4214 is not located in a foreground screen but is physically
associated with the warm ionized gas.

\end{abstract}


\section{INTRODUCTION}

	\objectname{NGC 4214} is an emission-line selected dwarf 
galaxy. It was chosen for observation based on a study of a 
complete sample of galaxies from the Center for
Astrophysics redshift survey (CfaRS; \citealt{Huchetal83b}) presented by 
\citet{Burg87} \footnote{The CfaRS consisted of 2400 galaxies taken from the 
original Zwicky Catalog to a limiting magnitude of $M_B^0 = 14.5$.}. The Burg 
study, which included emission line data for the galaxy, noted that NGC 4214
possesses optical emission line ratios (log \oiiir\ / \hb\ $> 0.2$ and 
log \niir\ / \ha\ $< -0.6$) that are similar to those of giant \ion{H}{2}
regions and other starbursts \citep{Baldetal81}. Its metallicity has been
measured by \citet{KobuSkil96}, who find 12 + log(O/H) = $8.20-8.36$. The 
distinguishing qualities of this object are: (1) It is among a set of the most 
highly excited photoionized galaxies of the CfaRS. (2) It is listed as a 
Wolf-Rayet galaxy \citep{Schaetal99b}, 
exhibiting broad \ion{He}{2} 4686 \AA\ emission in the heart of its brightest 
clusters \citep{SargFili91,MasHKunt91b}. (3) It is very bright at UV continuum 
wavelengths \citep{Faneetal97}. NGC 4214 has a $M_B^0 \approx -18.8$ if located
at a distance of 4.1 Mpc \citep{Leitetal96}, the value which will be used in
this article, even though \citet{Hoppetal00} suggest that it could be somewhat
closer.

	The brightness of NGC 4214 in the UV may be related to its 
apparently low dust content. The dust mass, inferred from its far-IR continuum 
luminosity, L$_{\rm IR}$ ($\sim 1.1\cdot 10^9$ L$_{\odot}$, 
\citealt{Throetal88}), is $1.8\cdot 10^5$ M$_{\odot}$, much smaller than dust 
masses typically found in Ultraluminous Infrared Galaxies and even among other 
dwarf starbursts. Furthermore, the ratio log(L$_{\rm IR}$/L$_{\rm UV}$) is 
$\sim 0.82$, much smaller than is found for most of the UV bright starbursts 
observed with IUE \citep{Hecketal98}. This indicates that most of the observed 
UV emission is not reprocessed into far-IR emission via dust absorption on 
global scales. Geometry may play a large role, since NGC 4214 has a nearly face 
on orientation on the sky \citep{Alls79} and apparently a thin disk 
(\citealt{Maizetal99a}, hereafter, MMTM). However, there are subgalactic 
variations of dust extinction which complicate our analysis. \citet{Maizetal98} 
(hereafter, MMMVC) report irregular variations of $E(B-V)$ on scales of
order $\sim$ 0.1-0.6 arcseconds, based on their spatial map of Balmer line 
ratios in the inner regions of the galaxy. 

\section{OBSERVATIONS AND DATA REDUCTION} 

\subsection{HST/WFPC2 imaging}

\begin{figure}
\centerline{\includegraphics*[angle=270,width=\linewidth]{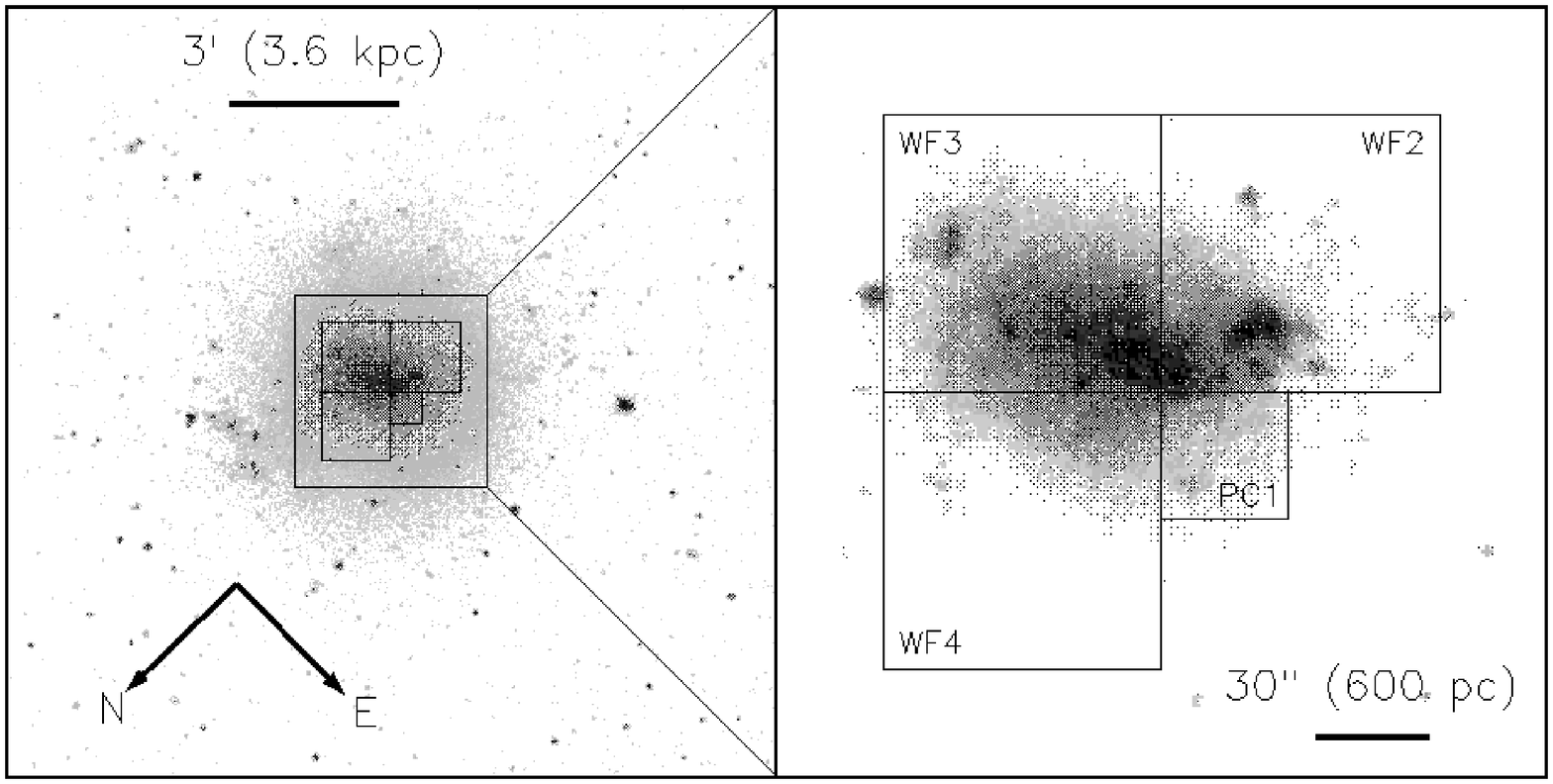}}
\caption{Digitized Sky Survey image of NGC~4214 with the WFPC2 field 
superimposed on it. The contrast in the large scale image on the left is 
stretched to show the low brightness disk of the galaxy while the one in the
blow-up on the right yields a more realistic perception.}
\label{dssimage}
\end{figure}

 	We obtained deep, high resolution, multiwavelength 
imaging of NGC 4214 with the WFPC2 instrument aboard HST (prop. ID 6569) on 
1997 July 22. In Figure~\ref{dssimage} we 
show a graphic representation of the WFPC2 field superimposed on a 
$13\farcm 6\times 13\farcm 6$ greyscale reproduction of the Digitized Sky Survey
Image\footnote{Based on photographic data of the National Geographic Society -- 
Palomar Observatory Sky Survey (NGS-POSS) obtained using the Oschin Telescope 
on Palomar Mountain. The NGS-POSS was funded by a grant from the National 
Geographic Society to the California Institute of Technology. The plates were 
processed into the present compressed digital form with their permission. The 
Digitized Sky Survey was produced at the Space Telescope Science Institute 
under US Government grant NAG W-2166.}. The equatorial coordinates of the 
intersection point of the four camera fields are 
$\alpha = 12^{\rm h}15^{\rm m}40\fs 53$,
$\delta = 36\arcdeg 19\arcmin 37\farcs 6$ (J2000). This location was chosen 
in order to minimize Charge Transfer Efficiency (CTE) effects, since in this 
way no area of interest is separated from its collecting point by low signal 
areas \citep{Whitetal99}. At a distance of
4.1 Mpc, $1\arcsec = 20$ pc, so that one WF pixel corresponds to 2.0 pc
and one PC pixel to 0.91 pc. Spatial resolution on these physical scales is 
within a factor of $2-4$ of the best resolution that can be achieved from the 
ground for the LMC. Thus, the high spatial resolution, high sensitivity, wide 
dynamical range and faint magnitude limit (roughly 3 magnitudes fainter than
typical ground-based limits, $\sim$ 26 in V) of WFPC2 permits one to identify 
regions structurally similar to those in the LMC (e.g.
\objectname{30 Doradus}), with ground-based resolution quality, for 
objects at nearly 80 times the distance to the LMC. Throughout the remainder of 
this paper, all conversions between pixels and arcseconds/physical scales 
are based on the plate scale of the WF chips, where most of NGC 4214 was imaged. 

	We summarize the WFPC2 data set analyzed for this study in 
Table~\ref{wfpc2obs}. A set of 15 HST/WFPC2 exposures was collected with 2
narrow band and 4 continuum filters. The continuum exposures consisted of two 
long ones and a short one. The short exposure was obtained in order to correct 
for the possible saturation at the location of strong point sources 
(unfortunately, the short F702W exposure was lost). 
The gain on the WF chips was 7$e^{-}$ $DN^{-1}$. Due to the limited
observing time and the desire to carefully remove cosmic ray events, we did not 
apply dithering techniques during the observing period, commonly used to 
partially compensate for the undersampled nature of the Point Spread Function 
(PSF). 

	The entire HST/WFPC2 data set was processed using the STScI WFPC2
pipeline. The images were then reduced and analyzed using various 
standard IRAF\footnote{IRAF (Image Reduction and Analysis Facility) is 
distributed by the National Optical Astronomy Observatories, which are operated 
by AURA, Inc., under cooperative agreement with the National Science 
Foundation.} packages as well as IDL procedures developed specially for this
purpose. Detailed photometric analysis of the continuum images will be 
presented in a later discussion (Paper II), which will compare some of the 
results of the resolved stellar populations of NGC 4214 with the integrated 
line emission measurements made in this study. Here we will use the continuum 
images only for continuum subtraction of the narrow band images and for 
integrated photometry at the wavelength of \ha.

	We used our multiple exposures to account for cosmic rays and we 
eliminated hot pixels and other defects. Each chip was flux calibrated using the
corresponding PHOTFLAM keywords, the sky (or zero point offset) was subtracted,
and we finally coadded the images to produce deep mosaics. 

	The analysis of the mosaics revealed that all the F702W, F555W, and
F336W exposures were saturated at the location of knot I-As (the knot
nomenclature is explained in the next section). Also, the two 
F702W exposures were saturated at knot IIIs (the nucleus of the galaxy). To 
eliminate the effect in the F555W and F336W exposures, we used archival WFPC2 
images from proposal 6716 (P.I.: T. P. Stecher) in which knot I-As was detected 
in the PC chip instead of in the WF3 one. The smaller pixel size allowed knot 
I-As not to be saturated in the F336W filter and to suffer only from slight 
saturation in the F555W, which was easily corrected. We used those images to 
substitute the saturated pixels and their neighbors. Unfortunately, no F702W 
images were available to correct for the saturation at knots I-As and IIIs.
Therefore, in that case we could only substitute the affected pixels by a 
linear combination of the F555W and F814W images, a procedure that introduces a
small uncertainty in the \ha\ continuum for those knots, the relevant quantity 
which will be used in this paper (see below).

	To eliminate the contribution of the continuum to the narrow band 
images and that of line emission to the continuum F555W and F702W images, we 
developed the following procedure. First, we assumed that the only nebular lines
contributing to the continuum images were \ha, \oiii, and \hb, with fixed
ratios \oiiir\ / \oiiil\ = \ha\ / \hb\ = 3.0, and we neglected other lines such
as \nii\ and \sii. Second, we approximated the spectral continuum flux 
at 5007 \AA, $F$($\lambda$5007), as the average spectral continuum flux
for the F555W filter, $F$(V), and the one at \ha,
$F$($\lambda$6563), by a linear combination of $F$(V) and $F$(R) (the 
average spectral continuum flux for the F702W filter), with 
coefficients determined from the STSDAS package SYNPHOT. Finally, we used 
SYNPHOT again to produce the following linear system:

\begin{equation}
\begin{array}{lcccccccc}
F(\rm{702W}) & = & 1.00\; F(\rm{R})       &   &                        & + & 
                   0.99\; F(\ha)    / w_1 &   &                        \\
F(\rm{656N}) & = & 0.76\; F(\rm{R})       & + & 0.24\; F(\rm{V})       & + & 
                   1.00\; F(\ha)    / w_2 &   &                        \\
F(\rm{555W}) & = &                        &   & 1.00\; F(\rm{V})       & + & 
                   0.36\; F(\ha)    / w_3 & + & 1.00\; F(\oiiir) / w_3 \\
F(\rm{502N}) & = &                        &   & 1.00\; F(\rm{V})       &   &
                                          & + & 1.00\; F(\oiiir) / w_4 \\
\end{array}
\label{linear1}
\end{equation}

\noindent Here, $F$(702W), $F$(656N), $F$(555W), and $F$(502N) are the PHOTFLAM 
calibrated measured spectral fluxes\footnote{Note that the standard WFPC2 flux 
calibration produces spectral fluxes even for narrow band filters.}, $F$(\ha) 
and $F$(\oiiir) are the true line fluxes and $w_1\ldots w_4$ are the effective 
widths of each filter (in Angstroms). We solved the linear system and dropped 
the smaller terms to obtain:

\begin{equation}
\begin{array}{lcrclcl}
F(\rm{R}) & = &          1.014\; F(\rm{702W}) & - & 0.018\; F(\rm{656N}) 
                                              &   &                       \\
F(\rm{V}) & = &          1.027\; F(\rm{555W}) & - & 0.025\; F(\rm{502N}) 
                                              & - & 0.006\; F(\rm{656N})  \\
F(\ha)    & = & 29.9\; ( 1.017\; F(\rm{656N}) & - & 0.773\; F(\rm{702W}) 
                                              & - & 0.250\; F(\rm{555W})) \\
F(\oiiir) & = & 35.6\; ( 1.027\; F(\rm{502N}) & - & 1.025\; F(\rm{555W})) 
                                              &   &                       \\
\end{array}
\label{linear2}
\end{equation}

	As is always the case when two measured quantities are subtracted,
the final relative uncertainty is largest for the case when the result is much 
less than the terms which are being subtracted. That would be the situation
when we want to calculate line fluxes where only stars are present or continuum
fluxes where there is mostly gas. The first case may apply to some small
apertures around isolated star clusters. However, given that most of the 
star forming episodes in NGC 4214 are quite recent (with large values of \wha\ 
and \woiiir), the corrections for the line fluxes (and the corresponding 
increase in the relative uncertainties) will always be small for large 
apertures and the largest uncertainty will always be the absolute photometric
calibration of the narrow band filters. The second case becomes important when 
the equivalent width of the line is comparable to the effective 
width of the filter, which occurs only in some very extreme situations for 
\ha\ or \oiiir\ (as will be shown later). Even then, since the estimated 
uncertainty in the coefficients which appear in the subtracting terms in 
Eq.~\ref{linear2} is not larger than 15\%, the final uncertainty will also be
smaller than that. That number should be compared with the error that occurs
if no correction is applied at all: since up to 50\% of the flux registered 
with the F555W or F814W filters may be actually coming from emission line 
photons, we may overestimate the continuum flux by a factor of two.

\subsection{VLA observations}

	NGC 4214 was observed with the NRAO Very Large Array (VLA) on 1988 
April 24 in the C configuration at 6 cm (4.885 GHz) and on 1989 March 16 in the 
B configuration at 20 cm (1.465 GHz). These wavelength - configuration 
combinations each result in synthesized beam diameters of approximately 4 
arcseconds. On-target integration times were 855 seconds at 6 cm and 700 
seconds at 20 cm. The integrations each were divided into two segments so that 
a larger range in hour angle (and a better sampling of the UV plane) might be 
achieved.

	The observations were edited to exclude bad antennas and samples. 3C286 
was used as a primary flux calibrator for both observations and phase 
calibrators were selected from the VLA recommended sample. The data were 
converted to maps using the standard AIPS software in 1989/90 with a 1 
arcsecond pixel size. The MX task was used to map and CLEAN the uv data. The 
resulting maps have rms noise levels of approximately 0.08 mJy at 6 cm and 
0.13 mJy at 20 cm.

\section{THE STRUCTURES OF THE IONIZED GAS}

\subsection{Narrow band morphology}

\subsubsection{Global structure}

\begin{figure}
\centerline{\includegraphics*[width=\linewidth]{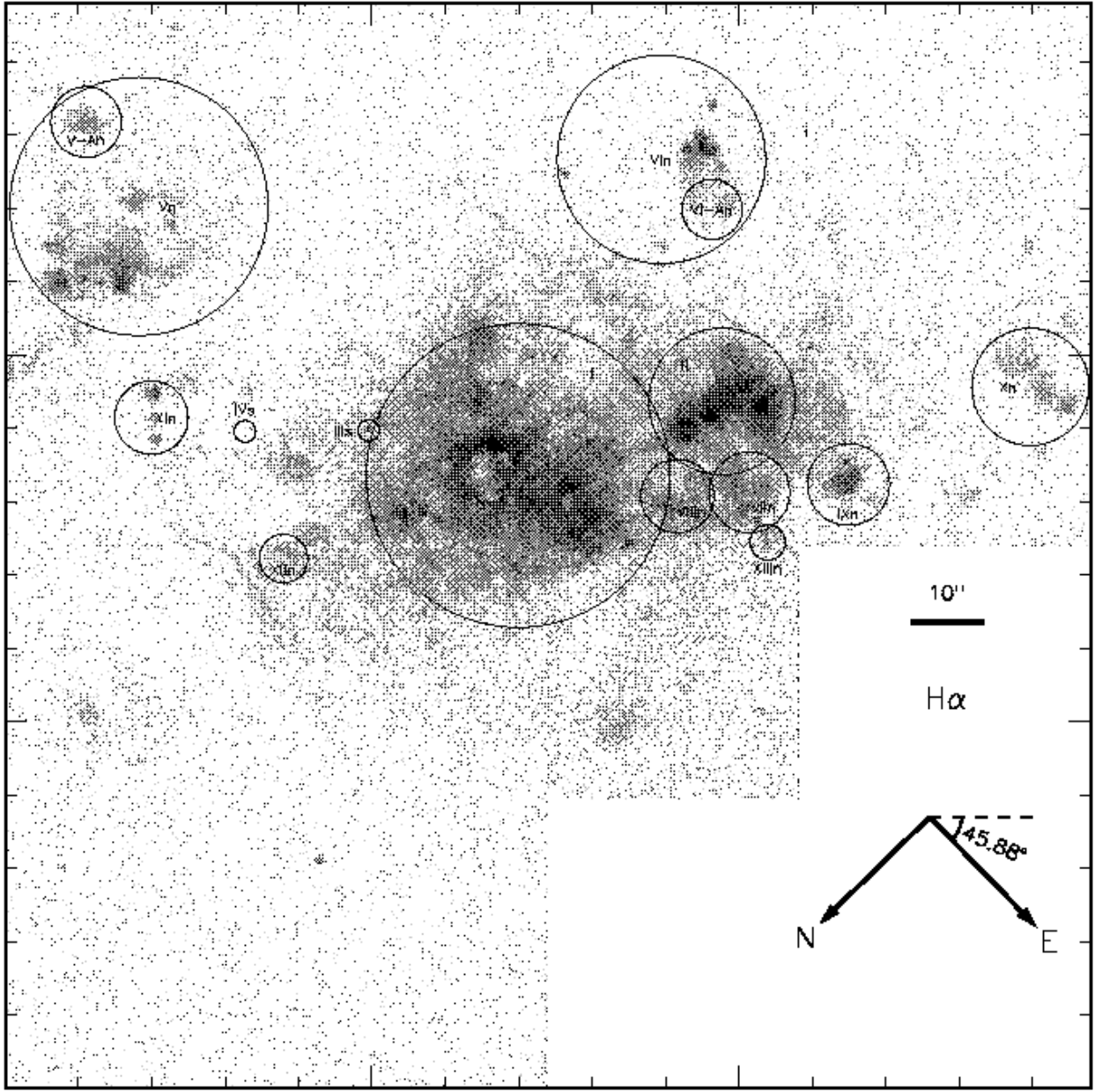}}
\caption{Logarithmic intensity scale, continuum subtracted \ha\ WFPC2 image of 
NGC~4214. We show and label the unit nomenclature and corresponding circular 
apertures used in this work which do not correspond to subdivisions of the two 
main \ion{H}{2} complexes, I and II. Those are shown in Fig.~\ref{oiiirhalpha}. 
Tick marks are shown for every hundredth pixel.}
\label{halpha}
\end{figure}

\begin{figure}
\centerline{\includegraphics*[width=\linewidth]{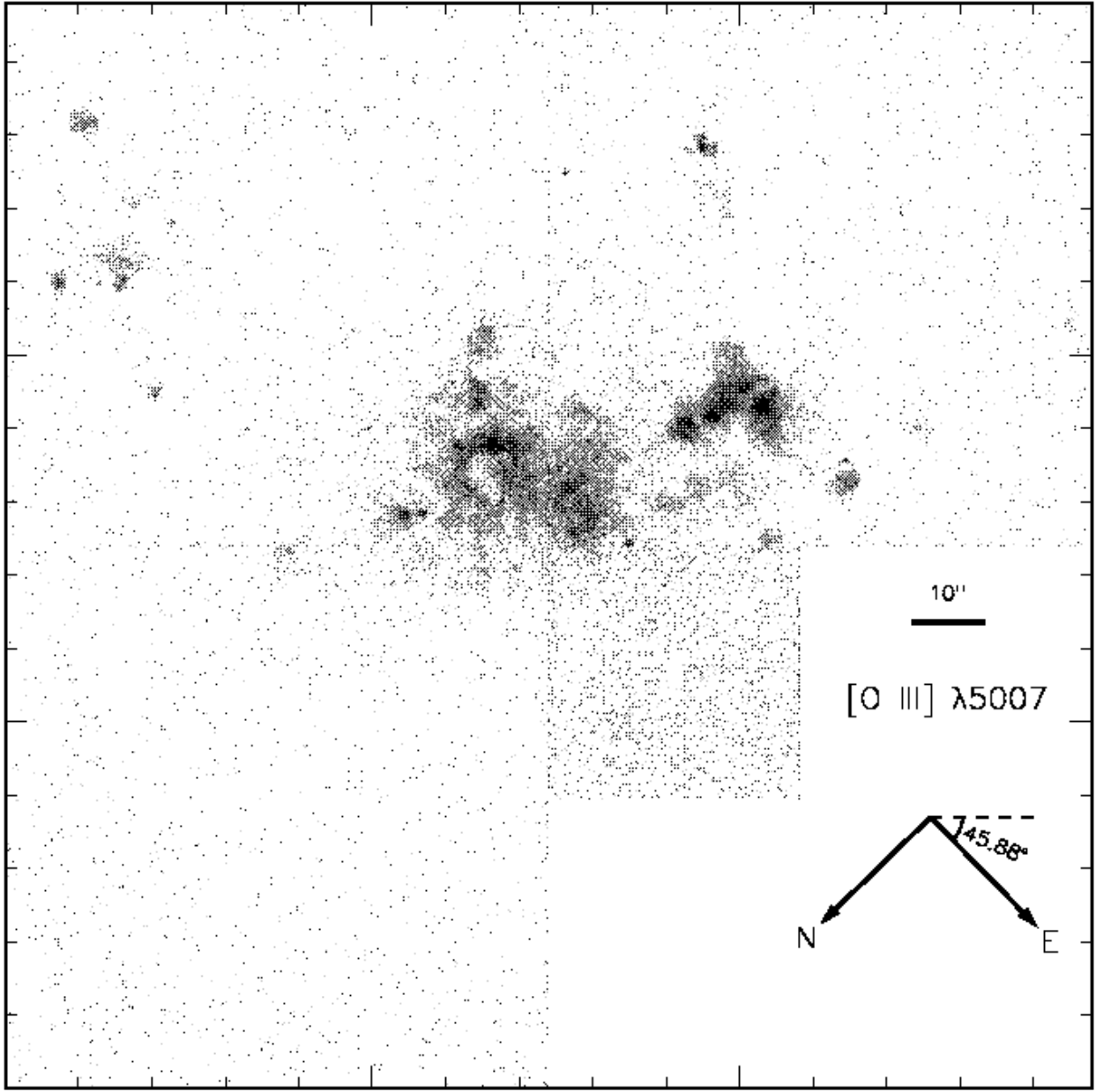}}
\caption{Logarithmic intensity scale, continuum subtracted \oiiir\ WFPC2 image 
of NGC~4214. Tick marks are shown for every hundredth pixel.} 
\label{oiiir}
\end{figure}

\begin{figure}
\centerline{\includegraphics*[width=\linewidth]{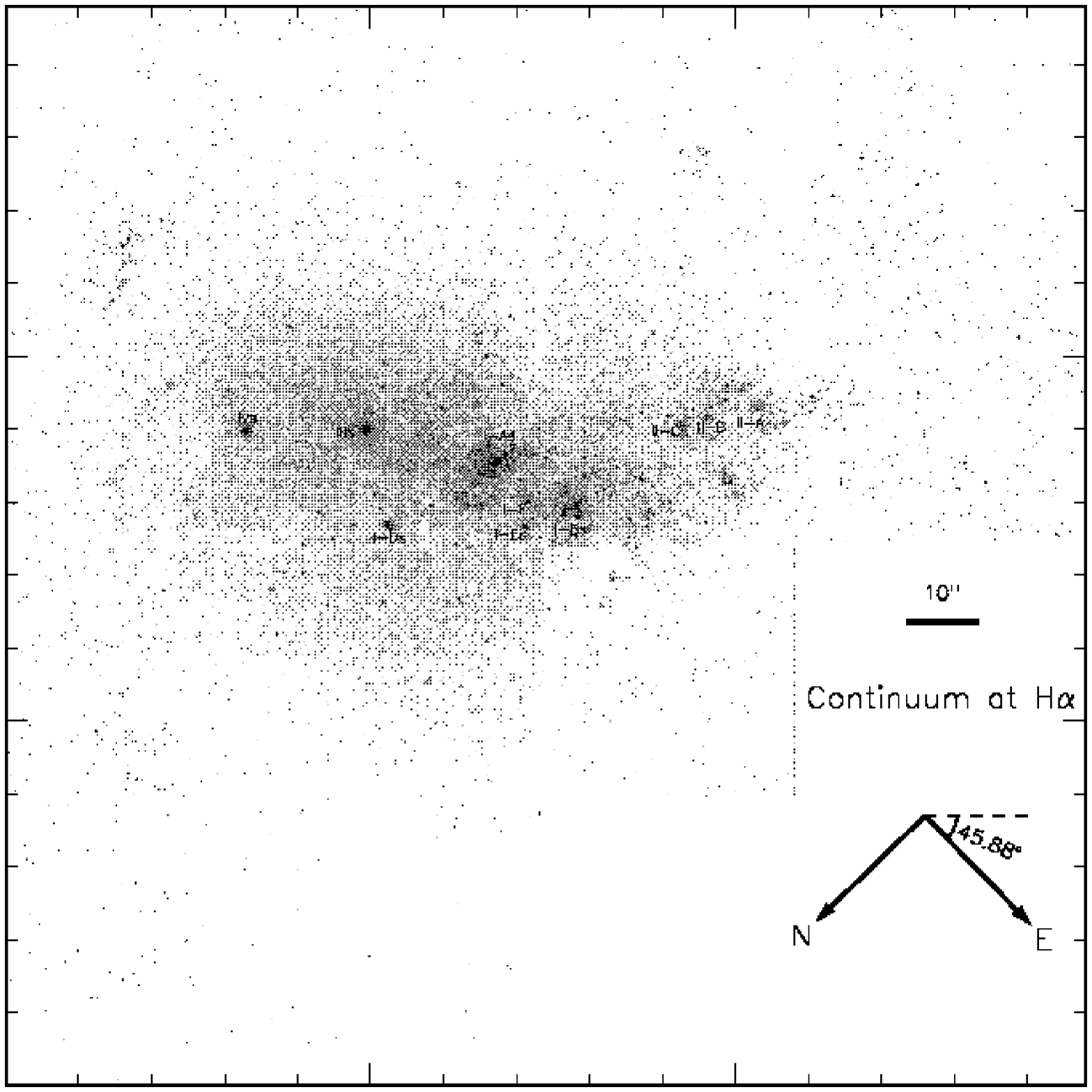}}
\caption{Logarithmic intensity scale \ha\ continuum WFPC2 image of
NGC~4214. This image is a linear combination of line emission subtracted V and R
images. We label the units which correspond to the main continuum knots. Ticks 
marks are shown for every hundredth pixel.}
\label{chalpha}
\end{figure}

\begin{figure}
\centerline{\includegraphics*[width=\linewidth]{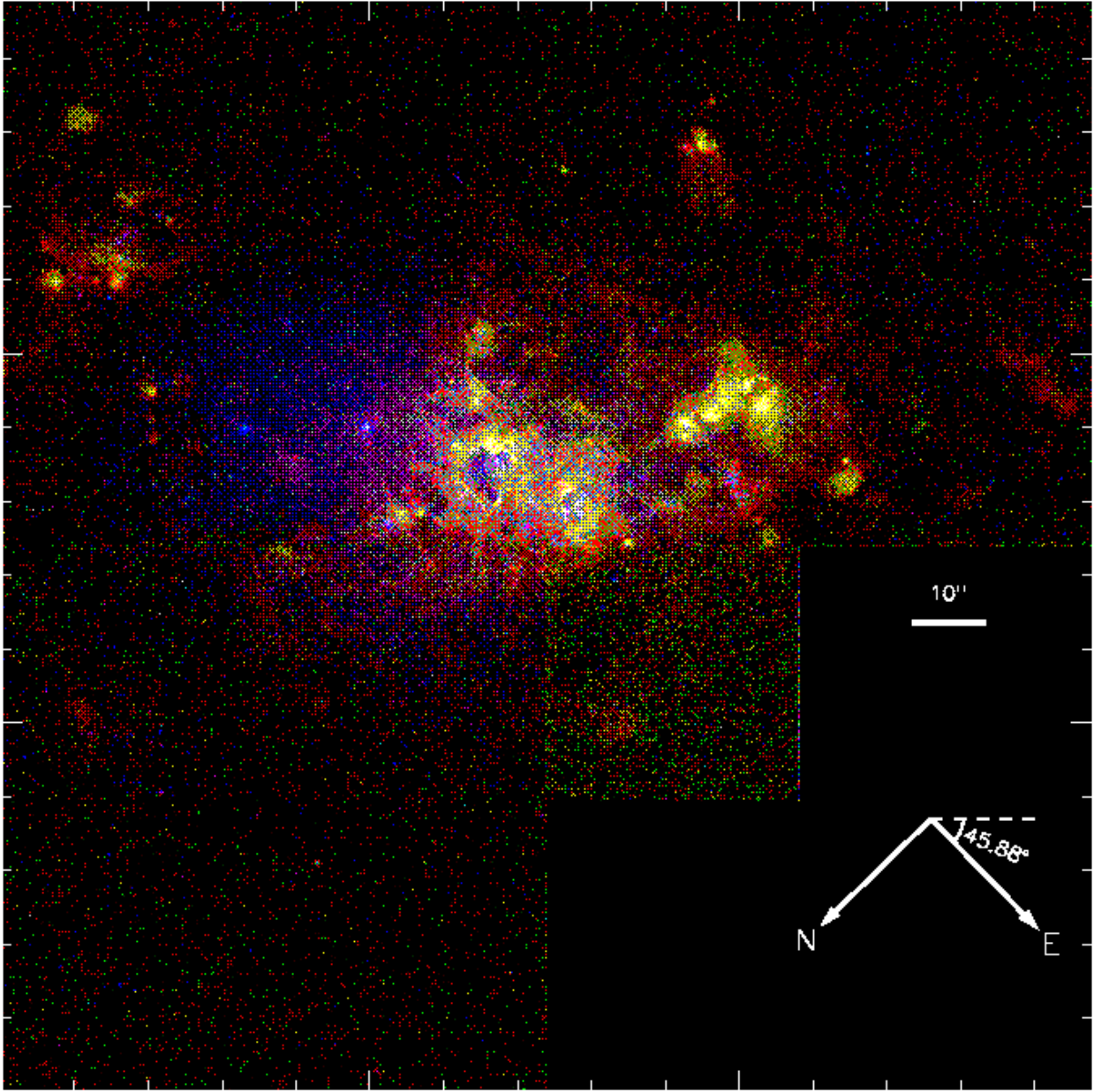}}
\caption{WFPC2 color mosaic of NGC~4214. Red corresponds to \ha\ line emission, 
green to \oiiir\ line emission, and blue to the continuum at \ha. The intensity
scale for each filter is logarithmic. The line emission filters have had the
continuum contribution subtracted. Tick marks are shown for every hundredth
pixel.}
\label{3cmosaic}
\end{figure}

	We show in Figs.~\ref{halpha}~and~\ref{oiiir} the two WFPC2 narrow band
images (\ha\ and \oiiir) on a logarithmic scale after continuum subtraction. We 
also show in Fig.~\ref{chalpha} the \ha\ continuum ($\lambda$6563) image 
produced as described in the previous section. The three images are combined 
into a three color mosaic in Fig.~\ref{3cmosaic}. 

	The narrow band morphology shows three differentiated components in
NGC~4214:

\begin{enumerate}
\item Two large \ion{H}{2} complexes, known in the literature as NGC 4214-I (or
NW complex) and NGC 4214-II (or SE complex), located near the center of the
field.
\item A number of isolated fainter knots scattered throughout the field,
especially in the upper half (SW) of the images.
\item Extended, structurally amorphous Diffuse Interstellar Gas (DIG) 
surrounding the two main complexes and some of the isolated knots.
\end{enumerate}

	The ISM exhibits a wide range of structural features of various
sizes, surface brightnesses, and luminosities. There are bright, compact knots;
rings and incomplete shells roughly $3\arcsec - 7\arcsec$ ($\sim 60 - 140$ pc)
in diameter; and fainter, wipsy filaments and arcs, whose dimensions 
range between roughly 5\arcsec\ and 20\arcsec\ ($\sim 100 - 400$ pc) along their
longest dimension.  Globally, the low surface brightness gas is patchy 
in the extreme outer parts while somewhat elliptical in the inner 
regions. Most of the high surface brightness features reside within the very 
inner regions, without any noticeable symmetric order. Although there are a few 
cases of bright outlying regions of isolated knots, the structure of the narrow 
band emission of NGC 4214 becomes increasingly more extended and filamentary 
and of lower surface brightness as one goes from the inner to the outer 
regions. It is, thus, natural to think of NGC 4214 as a centrally concentrated 
region of two large \ion{H}{2} complexes within 15$\arcsec$ or 300 pc of the 
center of the WFPC2 field and some fainter isolated regions. Both types 
of star forming units are superimposed on a more extended DIG ($> 30\arcsec$
or 600 pc), which is visible only in \ha\ but not in the high excitation 
\oiiir\ line. This type of global morphology was well noted in the past 
\citep{Hodg69,Hodg75,Hunt82}. 

	The morphology apparent in the radio continuum images of 
Fig.~\ref{radio} is different at the two wavelengths. The 6 cm image shows a
number of peaks which coincide with the brightest \ha\ peaks, indicating that
most of the flux is produced by thermal emission in the \ion{H}{2} regions. The
relative peak intensity among knots is not the same as in the \ha\ images due 
to different amounts of extinction. On the other
hand, the 20 cm image shows a morphology which is not so well correlated with
\ha. Even though some knots appear at the same location of the \ha\ or 6 cm
ones, there are several others with non-existent or weak counterparts at those
wavelengths. In those cases most of the emission has a non-thermal origin with a
more negative spectral index which causes the corresponding 6 cm emission to be
weak.

	The morphology in the \ha\ continuum image of Fig.~\ref{chalpha} is 
quite different from the narrow band one. There we see a number of clusters
dispersed parallel to the horizontal axis of the image superimposed on a 
well-defined ellipsoidal background with its major axis inclined 15\arcdeg\ with
respect to that direction. A large number of individual bright stars are 
scattered across the whole WFPC2
field. Some of the clusters are associated with narrow band structures while 
others show no such correspondence. The ellipsoidal unresolved background 
corresponds to the old population in the bar of NGC 4214 \citep{Faneetal97} and 
the probable nucleus of the galaxy is indicated as IIIs (the nomenclature is 
explained below). The two main complexes are not centered with respect to the
ellipsoid but displaced toward the E-SE, with NGC 4214-II located further away
from the nucleus than NGC 4214-I.

	The individual \ha\ knots have been identified by several authors, most
recently by MMMVC, who provide in their Table 2 a list of 
equivalences with some of the older nomenclatures. Here we present in
Table~\ref{apertures} a more standardized nomenclature with 13 different units
and the corresponding equivalences to the one defined by MMMVC using lower 
resolution and coverage\footnote{Those authors used a classification scheme in
which numbered knots corresponded to \ha\ maxima while lettered ones
corresponded to continuum maxima.}. We decided to use a classification which 
was as consistent as possible with previous work and which also identified the
different physical units. However, we note that with the present data it is 
possible that the assignment of physical character to some of the units may be 
uncertain and we also note that even in the high resolution HST images, there is
some subjectivity in separating individual \ion{H}{2} regions. Thus, it
is possible that ``well resolved'' clumps may actually be a collection of a few 
smaller individual objects.

	The first two units, NGC 4214-I and NGC 4214-II, 
correspond to the main \ion{H}{2} complexes in the center of the WFPC2 field
(previously, NW and SE complexes). NGC 4214-III and NGC 4214-IV are two large 
compact continuum sources to the left (NW). Then, NGC 4214-V to NGC 4214-XIII 
are progressively fainter \ha\ structures. When appropriate, substructures 
within each unit are assigned a letter and named I-A, I-B and so on. If even
smaller \ha\ knots are found, they are assigned a number (and named I-A1,
I-A2...). Finally, a letter n is added to the name in each entry in 
Table~\ref{apertures} if the aperture is defined with respect to the narrow
line image (because it is a pure nebular knot or because there is no obvious
large stellar cluster associated with it) or an s if it is defined with respect
to the continuum image (because it is a stellar cluster with weak or no nebular
emission around it or because it encompasses the location of the cluster in a 
unit with extended nebular emission). No letter is added if the aperture
includes one or several large stellar clusters and the nebular emission around
them. This nomenclature is better explained with a few examples: I-As is a
stellar cluster in NGC 4214-I with weak nebular emission at its position and
surrounded by several mostly nebular features I-A1n, IA2n\ldots, with I-A 
being used to refer to the whole region; II-A is a stellar cluster with intense
cospatial nebular emission (thus, with no need to define II-As or II-An); 
NGC 4214-IIIs is a stellar cluster with weak nebular emission if any. The 
position of each of the units and subunits is indicated in Figs.~\ref{halpha}, 
\ref{chalpha}, and \ref{oiiirhalpha}, as appropriate.

	We now analyze the different units in NGC~4214. For the rest of the
paper, except where noted, ages are obtained from MMMVC, who use a combination 
of criteria (\whb, \wwr, \lwr/\lhb, and $T_{\rm eff}$) to determine them.

\subsubsection{NGC 4214-I}

	NGC 4214-I is the largest \ion{H}{2} complex in the galaxy and also the
one with the most intricate morphology. It includes several star clusters and
has a complex \ha\ structure dominated by the presence of two cavities or
intensity minima.

	I-As is a massive young ($3.0-3.5$ Myr) Super Star Cluster (SSC) 
which was previously 
studied using HST FOC observations by \citet{Leitetal96}. Those authors find 
that the SSC core has a very small size (diameter $\lesssim$~5~pc) and contains
several hundred O stars. The core is in the middle of an extended star 
population which contains another several hundred O stars and extends up to a 
radius of $\sim$~70~pc. I-As is rich in Wolf-Rayet stars and shows wide \ha\
emission in its spectrum (\citealt{SargFili91},MMMVC). 

	Fig.~\ref{oiiirhalpha} shows that I-As is located in a heart-shaped 
\ha\ cavity, in between the center and the upper right corner. Our images 
show that the size of the cavity is $9\farcs 7 = 194$ pc in the NE-SW direction
and $8\farcs 0 = 160$ pc in the NW-SE direction. Little \ha\
emission is detected inside the cavity with the exception of that coming from
the SSC core and from the region immediately below and to its right (i.e.
toward the East). Some of the \ha\ emission detected from the SSC core must be
the wide component previously mentioned and some may be of nebular origin (the
core can also be seen in \oiiir\ in Fig.~\ref{oiiir}). The rest of the \ha\
emission may not
be real but caused by the incorrect continuum subtraction at that precise point
mentioned in the previous section. The weak \ha\ emission present in the
cavity shows three velocity components (MMTM), the most intense one
having a velocity similar to that of the surrounding gas and the other two being
displaced one toward the red and the other toward the blue. The secondary
components do not show the standard velocity ellipsoid corresponding to an
expanding bubble, which prompted MMTM to suggest that we are
witnessing here a bipolar flow that has punctured the plane of NGC 4214. The
red secondary component has a well defined intensity maximum in the spectra of
MMTM just below the core of the SSC. In our high resolution 
WFPC2 image (Fig.~\ref{oiiirhalpha}), we detect there the presence of two 
intrusions into the heart-shaped cavity which may be the source of that red 
component.

	The \ha\ emission surrounding I-As has a structure which shows
no spherical symmetry around the SSC. At the N border (lower left) of the
cavity, I-A2n is made out of an unresolved thin straight wall, which is probably
the ionization front produced by I-As, and a weak compact knot. The structure at
the E border, I-A4n, is distorted by the two intrusions previously mentioned 
but is otherwise similar. Both join almost at a right angle at the NE border of 
the cavity, helping define the heart shape. The S and W borders of the cavity
(I-A3n, I-A1n, and I-A5n) exhibit a more complex structure. Here the border is 
not so well defined, with two apparent linear structures at some points and with
some intense compact knots. As a matter of fact, the most intense \ha\ knot 
(in peak surface brightness) in all of NGC 4214-I is located here, at the 
smaller of the two apertures drawn in I-A1n in Fig.~\ref{oiiirhalpha}. 

	I-Bs is a $3.0-3.5$ Myr old Scaled OB Association (SOBA, 
\citealt{Hunt99}), a massive cluster which, as opposed to an SSC, does not show 
a marked central concentration. The number of O stars in I-Bs is only 
$\sim$ 30\% that of I-As (\citealt{Maiz99}, hereafter M99). Some of its stars 
are of Wolf-Rayet type \citep{SargFili91}. I-Bs is also located in the middle 
of an elongated \ha\ cavity, though not as clearly defined as the I-As one. 
However, the gas inside the cavity does show the signature of an expanding shell
(MMTM). Our images show that the cavity has a size of 
$8\farcs 5 = 170$ pc in the NE-SW direction and $3\farcs 0 = 60$ pc in the 
NW-SE direction. The \ha\ knots which surround the cavity seem to be formed
by a series of linear structures plus an intense knot (I-B1n) at its SW border.
In the \ha\ image the cavity appears to be broken between I-B1n and I-B2n. It is
just to the right (SE) of this apparent hole that MMTM detect an
expanding \ha\ bubble, which could be caused by the puncture of the cavity at
that point.

	Two smaller clusters can be seen between I-A and I-B: I-Es and I-F. I-Es
appears to have a quite compact structure, making it another SSC candidate
(though probably smaller than I-As). The \ha\ emission around it is not too
intense and appears to be part of the general NGC 4214-I background. Also, the 
continuum colors in the MMMVC data are significantly redder than for I-As or 
I-Bs, indicating a greater age ($\gtrsim$ 10 Myr). On the other hand, I-F must 
be a young cluster since there is \ha\ emission directly associated with it and 
its continuum colors are similar to I-As or I-Bs. 

	No large cluster is seen in the I-Cn aperture and the \ha\ emission
there is probably produced by a loose group of young stars located to the SW of
I-A. I-Ds is a SOBA somewhat smaller than I-Bs (M99). Much less \ha\ emission 
is observed around it in comparison to I-A or I-B, with only two (I-D1n and 
I-D2n) or possibly three compact \ha\ knots associated with it. I-Gn is a 
heavily reddened compact \ha\ knot (MMMVC) which falls quite close to the 
border between the PC and WF2 chips, hindering the study of its morphology with 
the present data.

\subsubsection{NGC 4214-II}

	NGC 4214-II is the second largest complex in the galaxy in size and
harbors the regions with the highest peak \ha\ intensity. Its morphology is 
quite different from that of NGC 4214-I. In the first place, there is no 
dominant SSC but several smaller SOBAs which are responsible for the ionization 
of the gas. Also, there are no large cavities present and the clusters are 
located very close to (or at the same position as) 
the most intensely \ha\ emitting knots
(MMMVC). Finally, no kinematic anomalies are detected here: All the
\ha\ profiles are gaussian at a resolution of 18 km s$^{-1}$, indicating that
expanding shells are non-existent or too weak in comparison with the main line
emission (MMTM).

	II-A and II-B are the two brightest \ha\ knots in NGC 4214-II. They are
both located on top of SOBAs which, at the resolution of WFPC2, appear to be
centered at the same positions. The centers of the knots suffer from little
extinction but the \ha\ emission in their surroundings is highly extincted
(MMMVC). A dust lane runs from S to N to the W of knot II-A, extending
up to knot II-B. Another one separates this last knot from II-C.

	The SOBA in II-B seems to be more compact than the one in II-A, which
extends toward the E. There, at a distance of $\sim 3\farcs 5$ ($\sim$70 pc) 
we find an incomplete \ha\ shell which is especially prominent in \oiiir.
We have reanalyzed the data of MMTM and found no kinematic 
anomalies in the \ha\ profile at the shell position.

	From the ground II-C appears to be a single, slightly extended knot and
here is clearly resolved into an incomplete shell divided into two fragments,
II-C1n toward the SW and II-C2n toward the NE. The diameter of the shell is
2\arcsec\ (40 pc) and the stellar cluster is located closer to the brightest of 
the two fragments, II-C1n. Despite the appearance of an
expanding broken shell, the kinematic \ha\ profile reveals a single \ha\
component (MMTM).

	II-Dn and II-En are two weaker \ha\ knots with relatively fewer stars at
their positions.

\subsubsection{Other units}

	NGC 4214-IIIs is a compact continuum source with weak associated \ha\ 
emission. It is probably an old SSC or the nucleus of the galaxy, as suggested 
by the study of \citet{Faneetal97} in the I band and the FUV. In the I band 
image of those authors, NGC 4214-IIIs appears at the center of the weak disk of 
the galaxy. The \ion{H}{1} observations of \citet{McIn98} also place the 
rotation center very close to NGC 4214-IIIs, favoring the nuclear 
interpretation.

	NGC 4214-IVs has a similar appearance to NGC 4214-IIIs and it also
appears in the I band image of \citet{Faneetal97}. However, it is dimmer
in the continuum images and here no \ha\ or \oiiir\ emission is detected at all 
except for extremely weak DIG close to the noise level which is probably 
unrelated to the continuum knot. NGC 4214-IVs is probably another old SSC or a 
background galaxy.

	The rest of the structures visible in the \ha\ image can be classified
as diffuse individual knots (VIIn, VIIIn, IXn, XIIn, XIIIn), series of knots
(Xn, XIn) or complex \ha\ knot + shell structures (Vn, VIn). Some areas of 
those knots appear as green in Fig.~\ref{3cmosaic} due to their high excitation.
Finally, some very weak additional structures can be seen in the \ha\ map due
to the use of a logarithmic intensity scale. They are only barely above the
noise level and are not visible in the \oiiir\ map. Therefore, little
information about them can be gained with the present data and they have not
been included in the aperture list.

\subsection{\oiiir\ / \ha\ maps}

	The ratio \oiiir\ / \hb\ is one of the excitation ratios used by
\citet{Baldetal81} and \citet{VeilOste87} to classify emission-line objects as a
function of the ionizing source (photons vs. shocks) and its spectrum. For
photoionized nebulae, \oiiir\ originates in a region smaller than
\hb. The reason is that the second ionization potential 
of oxygen is 35.11 eV, much higher than the first ionization potential of
hydrogen, 13.60 eV. The photons that can ionize \ion{O}{2} are absorbed close
to the source, producing an \oiiir\ + \hb\ inner emitting layer while in an 
outer layer only \hb\ is produced. The \oiiir\ / \hb\ ratio is a measure of the
hardness of the spectrum: The higher the fraction of high energy ionizing 
photons with respect to low energy ionizing ones, the larger the first layer
will be with respect to the second one and the larger the ratio. Unfortunately,
there are other factors which also enter in the equation to determine the final
outcome, the most important ones being the metallicity, the density, and the 
distance from the source to the beginning of the \oiiir\ emitting layer. 
Therefore, even though the \oiiir\ / \hb\ ratio can provide some useful 
information on the properties of the ionizing source, that information is 
limited unless additional data regarding the above mentioned parameters are
provided. 

	Here we present a map formed by the division of our continuum subtracted
\oiiir\ and \ha\ images (see Fig.~\ref{oiiirhalpha}). The data have been
smoothed with a 5 pixel box before the ratio was evaluated. The areas with low
signal to noise ratio have been left blank to eliminate confusion.
Note that the use of \ha\ instead of \hb\ in the denominator provides us with a 
better signal-to-noise ratio but is more heavily affected by extinction. 
Fortunately, we can distinguish between excitation and extinction effects using 
the \ha\ / \hb\ maps obtained by MMMVC and M99 using ground based data,
which are explained in more detail in the next section.

	The excitation ratio map shows very different patterns in the two main
\ion{H}{2} complexes. In NGC 4214-II there is a generally good correlation 
between \ha\ intensity and excitation, as previously noted by MMMVC. The
five units II-A to II-En appear as excitation maxima. The highest ratio is
reached in the brightest one, II-A, and it is not centered at the \ha\ peak but 
is located between that point and the broken shell to the E previously 
described. High \oiiir\ / \ha\ values are detected throughout most of the II-A 
aperture and also at the area adjacent to it toward the E. II-B also shows 
high excitation but in this case the maximum is located at the \ha\ peak. II-Dn 
and II-En can also be detected as maxima in the excitation maps while II-C1n 
and II-C2n are not so well defined.

	The structure of the excitation map in NGC 4214-I is more chaotic. The
most obvious maximum in the ratio around I-A appears at the location of the SSC
itself. However, as mentioned previously, the value of the ratio there is
affected by saturation and we will not consider it
here due to its questionable validity. The second obvious maximum is located
at the center of the \ha\ cavity, close to an area which appears in white due to
the low value of the signal-to-noise ratio. Again, this excitation maximum may
not be real and will not be analyzed here. Having said that, the most obvious
real \oiiir\ / \ha\ maximum in I-A is located at the brightest \ha\ peak, I-A1n.
However, the maximum is not as marked as for II-A or II-B. In its surrounding 
areas, I-A3n and I-A5n, the structures in the \ha\ image can be traced with some
difficulty and that task becomes even harder in the rest of I-A.

	The situation described for I-A becomes even more extreme in I-B. Here,
the most intense \ha\ knot, II-B1n not only is not a high excitation maximum 
but it sits at the tip of a low \oiiir\ / \ha\ finger which penetrates from the
left. Even though some dust lanes exist in the area, this structure is not due
to extinction since it can be seen in the low resolution maps of
MMMVC. In general, the correlation between \ha\ intensity and
excitation here is very weak.

	The highest excitation ratios in NGC 4214-I are found outside I-A and
I-B. I-C1n shows an excitation structure and values similar to those of II-A, 
with high values at the \ha\ peak and even higher values at a position nearby. 
Knot I-Hn shows an even higher \oiiir\ / \ha\ value and its surrounding areas 
toward the E are also of high excitation, even though their \ha\ intensity is 
quite low. This anomaly was already detected by MMMVC, who noted that it
appeared only in \oiiir\ / \hb\ but not in \niir\ / \ha\ or in \sii\ / \ha. At
this resolution, a N-S excitation ratio gradient which crosses most of NGC
4214-I can be observed.

	The rest of the structures seen in the \oiiir\ / \ha\ map generally 
follow the same scheme described for NGC 4214-II: high excitation values 
correspond to \ha\ peaks. We only note here that NGC 4214-XIIIn shows
anomalously high excitation values.

\section{INTEGRATED PHOTOMETRY}

\subsection{WFPC2 data}

	We have integrated the \oiiir, \ha, and \ha\ continuum fluxes for the
apertures in Table~\ref{apertures}. The results are shown in
Table~\ref{intfluxes}. We also show the corresponding values of the integrated
ratios, \oiiir\ / \ha\ and \wha\ in Table~\ref{intratios}. 

	Following \citet{Maizetal00a}, we present our results using three 
different extinction corrections: no extinction, foreground variable dust 
screen, and dust-gas mixed (the validity of these models is later analyzed in 
the discussion). No data are presented for the continuum at \ha\ 
using a mixed model due to the unphysical character of such an application 
(most of the continuum originates in stars which can have a spatial 
distribution along the line of sight quite different than that of the gas). For 
\wha\ several possibilities are considered: no extinction, only gas extincted 
(foreground or mixed models), and both gas and stars foreground extincted. The 
extinction data are taken from the MMMVC and M99 maps and have a 
variable resolution from $\approx 1\arcsec$ for the areas with the highest \ha\ 
intensities to $\approx 5\arcsec$ for low intensity ones. The data cover all 
of NGC 4214-II, most of NGC 4214-I, and other areas of interest. Outside the
coverage we have used an extinction typical of the DIG surrounding NGC 4214-I 
and II. Those integration areas where most of the flux is not included in the
M99 maps are marked with an asterisk in Table~\ref{apertures} to
indicate the uncertain character of the extinction correction there.

	Most of the \ha\ and \oiiir\ emission comes from NGC 4214-I and NGC
4214-II. Table~\ref{intfluxes} reveals the large uncertainty caused by the
distribution of dust with respect to gas. The correction due to extinction
applied to all of NGC 4214 more than doubles for both \ha\ and \oiiir\ when we
consider a mixed model instead of a foreground screen. In some cases it is so
large that the real values of the fluxes are extremely uncertain. Such is the
case for I-A1n, I-A3n, I-Gn, II-A and II-B. The most extreme case is I-Gn. 
There we are at the point of the extinction curve where the \ha\ / \hb\ ratio 
cannot give a good estimate of the real extinction if dust and gas are mixed 
\citep{Maizetal00a}.

	The \oiiir\ / \ha\ data in Table~\ref{intratios} reveal that, in
general, NGC 4214-II has a higher excitation than NGC 4214-I. An excitation 
gradient appears to be present in NGC 4214-II with high values to the right
(II-A) and low ones to the left (II-C). In I-A high excitation values are found
only in the compact knot in I-A1n and, to a lesser degree, in the one in I-A3n. 
The excitation in I-B is even lower than in I-A. In other regions
of NGC 4214-I we find high excitation in I-C1n, I-Gn, and I-Hn.

	Several isolated \ha\ knots exhibit stronger than expected excitation
values. We have already mentioned the cases of I-Hn and XIIIn, but the same is
also true for V-An, VI-An, and XIIn (the high value of NGC 4214-IVs is 
artificially produced by the low signal to noise ratio of the emission lines and
the incorrect continuum subtraction there). The most extreme cases are V-An and 
XIIIn, where the reddened value of \oiiir\ / \ha\ is higher than 1.5.

	The equivalent width of the Balmer lines can be used to estimate the age
of a star-forming knot (see, e.g. \citealt{CervMasH94}). However, there are
several problems that complicate the age estimation, namely, the choice of a
consistent aperture, the differential reddening, and the presence of an 
underlying old population (MMMVC). Here the two first problems can 
be easily solved while for the last one we can resort to the estimates of
MMMVC. To establish the correspondence we use the models of
\citet{Cervetal00} and \citet{Leitetal99} to plot in Figure~\ref{wha} the 
relation between age and \wha. The differences between the predictions are
caused mainly by the assumptions about the number of hydrogen-ionizing photons
which are absorbed by dust or escape (0\% in the \citet{Leitetal99} models and
30\% in the \citet{Cervetal00} ones), and in the use of different stellar 
atmospheres. 

	The expected values for \wha\ are $1\, 000-2\, 500$ \AA\ for very
young clusters (age $\lesssim 3$ Myr) and $500-1\, 000$ \AA\ for ages in the
range between 3 and $4-5$ Myr. This predicted decrease in \wha\ does not take 
into account the morphological evolution caused by mechanical energy 
deposition, since the fraction of effective to emitted ionizing photons is kept 
constant, as mentioned previously. Given that the surrounding gas is dispersed
in a timescale comparable to 5 Myr, in a realistic scenario the decrease would 
probably be even more pronounced.

	Values very close to or higher than $1\, 000$ \AA\ are found for all the
knots in NGC 4214-II except for II-C1n and II-C2n, indicating the very small 
age of this complex, as already pointed by MMMVC. The lower \wha\ values, 
broken shell appearance, and detection of nebular \ion{He}{2} emission in II-C
(probably caused by extreme WR stars, \citealt{SchaVacc98}) point toward a 
slightly greater age for this knot. The overall value of the \wha\
for NGC 4214-II is quite lower than $1\, 000$ \AA\ due to the existence of an
underlying population which is responsible for about $\sim 50\%$ of the \hb\
continuum (MMMVC). When that effect is taken into account, the 
measured estimated age for NGC 4214-II assuming a single burst model is
consistent with $2.5-3.0$ Myr.

	The \wha\ values for NGC 4214-I are in general lower than those of NGC
4214-II. When the underlying population is taken into account, an average age 
in the $3.0-4.0$ Myr range is the most plausible estimate. However, the two 
small apertures centered at knots I-A1n and I-A3n show values similar to those 
of the knots in NGC 4214-II, pointing toward a smaller age.

\subsection{Radio data}

	The lower resolution (4\arcsec\ in each band) and signal to noise of 
the radio data does not allow us to use the apertures of the analysis of the 
previous section. However, five distinct knots are visible in the 6 cm and/or 
20 cm maps and we have used those as reference points for the radio apertures. 
Two of the five apertures are located in NGC 4214-II, with each one of them 
roughly corresponding to its SE (II-AE) and NW (II-BCD) halves. The other three 
apertures are located in NGC 4214-I, roughly centered at the position of the 
I-A, I-F, and I-B WFPC2 apertures. We show in Table~\ref{radioap} the integrated
radio fluxes for the five apertures with the corresponding spectral index
$\alpha$.

\begin{figure}
\centerline{\includegraphics*[width=.45\linewidth]{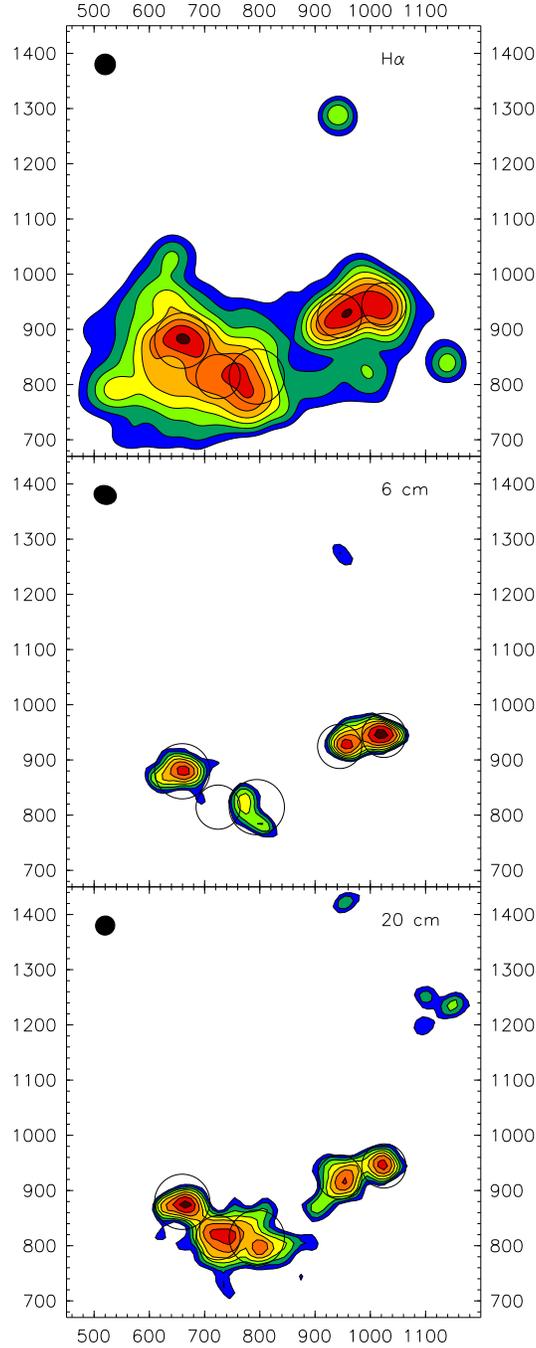}}
\caption{\ha\ (top), 6 cm (middle), and 20 cm (bottom) contour plots of 
NGC~4214. The \ha\ plot has been produced by degrading the resolution of the 
WFPC2 image to match approximately that of the other two, with the 
corresponding beam ellipsoids shown in the upper left corner in each case. The 
location of the five radio apertures is also shown. The eight levels are
logarithmically spaced, with the minimum and maximum placed at 
(15, 250) $10^{-16}$ erg s cm$^{-2}$ arcsec$^{-2}$ for
\ha, (15, 75) $10^{-6}$ Jy arcsec$^{-2}$ for 6 cm, and
(22.5, 90) $10^{-6}$ Jy arcsec$^{-2}$ for 20 cm, respectively. The orientation
is the same as in Fig.~\ref{halpha} and the coordinates are measured in pixels 
starting at the bottom left corner of the WFPC2 field.}
\label{radio}
\end{figure}

\begin{figure}
\centerline{\includegraphics*[width=.98\linewidth]{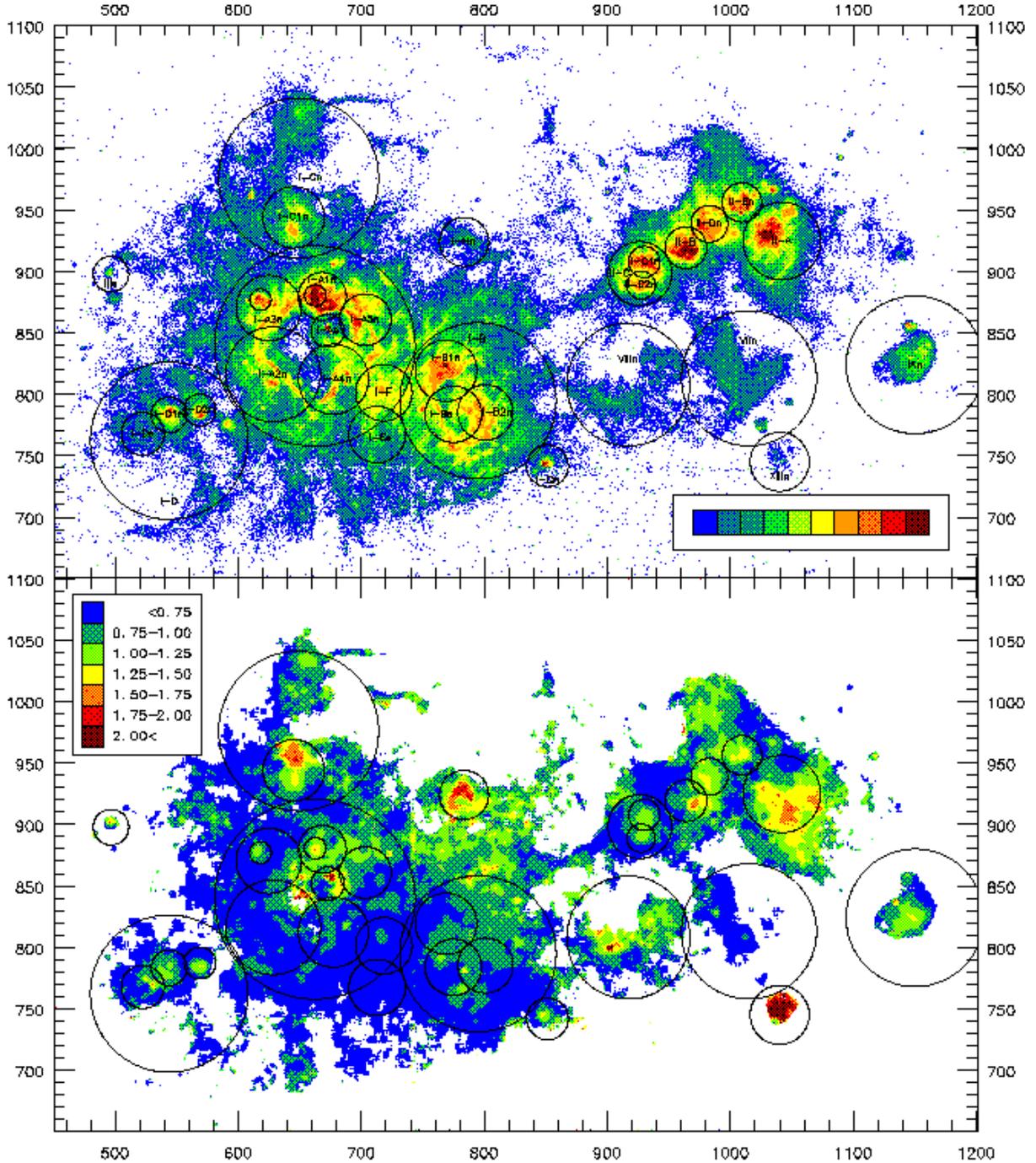}}
\caption{Top: Central area of the continuum subtracted \ha\ WFPC2 image. The
intensity scale is logarithmic between 20 and 1000 $10^{-16}$ erg s$^{-1}$ 
cm$^{-2}$ arcsec$^{-2}$. The apertures in this area are shown, with the 
exception of I and II (see Fig.~\ref{halpha}). All shown apertures are labelled,
with the exception of the small I-A1n and I-A3n ones. The orientation is the 
same as in Fig.~\ref{halpha} and the coordinates are measured in pixels 
starting at the bottom left corner of the WFPC2 field. Bottom: \oiiir\ / \ha\ 
map of the same area as the top part. The intensities were smoothed with a 5 
pixel box before evaluating the ratio and the areas left in white have \ha\ or 
\oiiir\ intensities less than 15 $10^{-16}$ erg s$^{-1}$ cm$^{-2}$ 
arcsec$^{-2}$. The same apertures are shown again to guide the eye.}
\label{oiiirhalpha}
\end{figure}

\begin{figure}
\centerline{\includegraphics*[width=.95\linewidth]{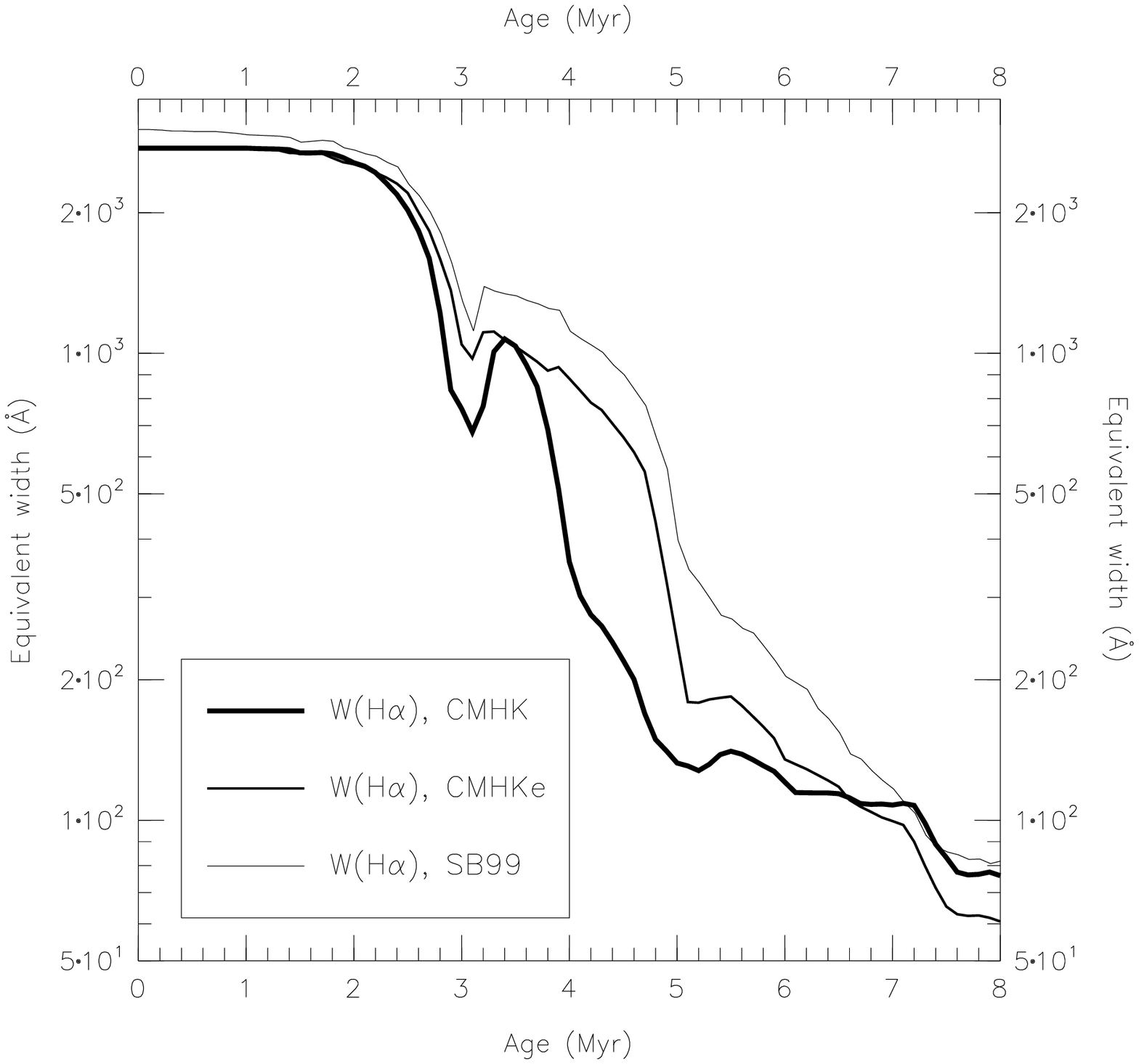}}
\caption{Predictions for \wha\ as a function of age from different evolutionary
synthesis models. The first two plots are from the models of \citet{Cervetal00}
(CMHK, online version available at  
{\tt http://www.laeff.esa.es/\~{}mcs/model}) while the third one is from the
Starburst 99 models of \citet{Leitetal99} (SB99, online version available at
{\tt http://www.stsci.edu/science/starburst99}). The first CMHK model uses
tracks with the standard mass-loss rate \citep{Schaetal93b}, while the other 
two models use tracks with an enhanced mass-loss rate for massive stars 
\citep{Meynetal94}. The atmospheric models used are \citet{Miha72}
and \citet{Kuru79} (CMHK standard mass-loss); \citet{Kuru79}, 
\citet{Schmetal92}, and \citet{Schaetal96a} (CMHK enhanced mass-loss); and 
\citet{Schmetal92} and \citet{Lejeetal97} (SB99). The two CMHK models assume 
that 30\% of the hydrogen ionizing photons are lost (absorbed by dust or 
escaped to the intergalactic medium) and use an upper mass limit of 120 
M$_\sun$. The SB99 models assume that no hydrogen ionizing photons are lost and 
use an upper mass limit of 100 M$_\sun$. All cases are for $Z=0.004$ and 
Salpeter IMF.}
\label{wha}
\end{figure}

	The spectral index of II-AE is almost exactly 0.1, the value expected
for a young ($<$ 3 Myr) star-forming knot with only thermal radio emission
\citep{MasH92}. II-AE is the brightest knot of all NGC 4214 at this resolution 
in the 6 cm image but only the third in the 20 cm map, which is consistent 
with such a high spectral index. The spectral index of II-BCD is not as high,
indicating that the radio emission is a mixture of thermal and non-thermal
emission. Looking at Fig.~\ref{radio}, we can see that the 6 cm emission appears
to originate from a point source toward the SE side of the aperture while the 
20 cm emission is more elongated toward the north. This is an indication that 
most of the thermal emission is coming from knots II-B and II-Dn while most of 
the non-thermal emission is originating in II-C.

	The spectral index of I-A is also quite high, indicating a predominantly
thermal origin of the emission. The position is consistent with an origin in
knot I-A1n and, to a lesser extent, I-A3n. However, a non-negligible fraction is
non-thermal and may be originating at knot I-A1n or at the nearby SSC. The
emission coming from the other two radio apertures has much lower spectral
indices. Even though some thermal emission may be originating from knots I-B1n
and I-B2n, undoubtely most of the radio flux detected there is non-thermal and
concentrated on the two knots clearly visible in the 20 cm image. The weaker of
the two non-thermal sources is centered close to the I-Bs cluster and extends
toward the SE. Its extent follows the approximate location of the superbubble
detected by MMTM which is expanding into the intercloud medium.

\begin{figure}
\centerline{\includegraphics*[width=\linewidth]{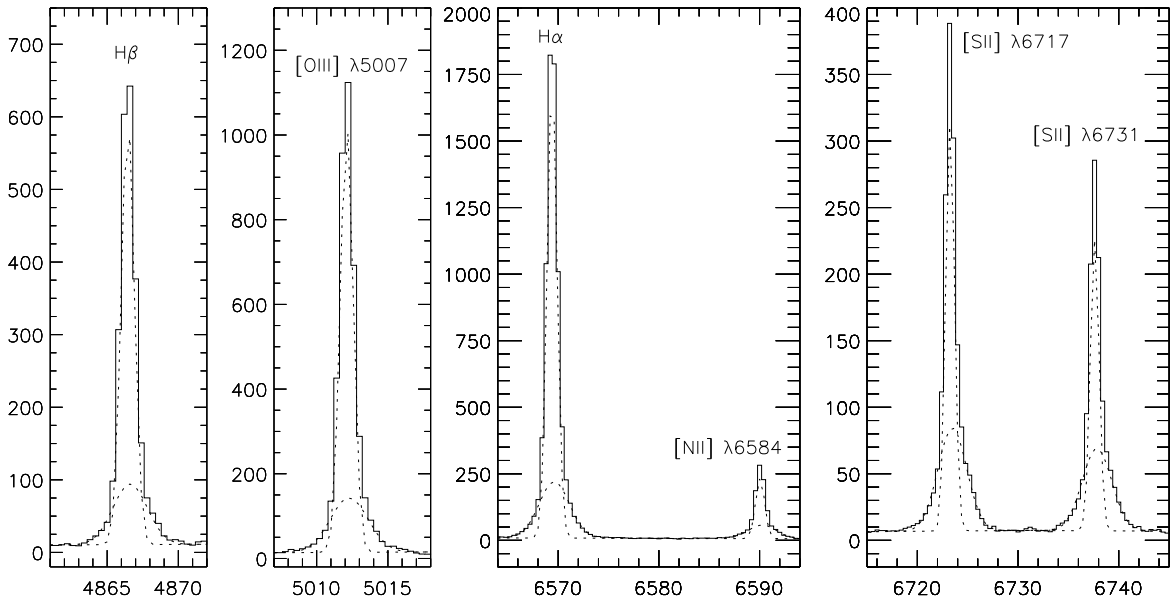}}
\caption{SNR optical spectrum extracted from the MMTM data. The fits for the
narrow and wide components are shown. The units are \AA\ and $10^{-17}$ erg 
s$^{-1}$ cm$^{-2}$ \AA$^{-1}$ arcsec$^{-2}$ for the vertical and horizontal 
axes, respectively.}
\label{snrspec}
\end{figure}

	The stronger non-thermal source is located within 2\arcsec\ of the 
stellar cluster in the WFPC2 aperture I-F but the resolution of the VLA data 
does not allow us to determine if they are spatially coincident. It has the
lowest spectral index in NGC 4214 and we identify it as a supernova remnant. We
have reanalyzed the data in MMMVC and MMTM and discovered clear signals of
the nature of this object. First, it appears as a well differentiated knot in
the \niir, \siil, and \siir\ maps but not in the \ha, \hb, \oiiir, or \hei\ ones
(see Fig. 3 of MMMVC). Second, a wide ($\sigma = 60-65$ km s$^{-1}$) component
appears in all of the analyzed emission lines, as is shown in
Fig.~\ref{snrspec}. The center of the wide component is displaced by $\approx$
10 km s$^{-1}$ toward the red with respect to the nebular emission. Finally, as
shown in Table~\ref{snrratios}, the wide component is the one responsible for
the localized high values of the \sii\ and \niir\ emission. Note that the
\oiiir\ / \hb\ ratio is not too different from the ambient nebular emission, 
which makes the SNR hard to detect with our WFPC2 data. The position of the wide
component in the long-slit data is also compatible with that of the cluster in 
I-F but, again, the spatial resolution is not good enough to make a certain
identification. We have also looked in the ROSAT archive and found an HRI image
of NGC~4214, but no strong x-ray source can be observed at this location.

	Some radio emission originating outside the two main \ion{H}{2}
complexes can be observed in the upper right quadrant of Fig.~\ref{radio}. One
knot appears to be thermal emission associated with NGC 4214-VIn. The knot 
located above it and the group to its right do not show up in the 6 cm image 
and are likely non-thermal sources. Thermal emission from knots VIIn, VIIIn, or 
IXn is not detected.

	The ratio of thermal radio emission to \ha\ can be used to calculate the
true optical depth $\tau_{\rm rad}$ at \ha\ caused by dust (see, e.g. 
\citealt{ChurGoss99}):

\begin{equation}
\tau_{\rm rad} = 
\ln\left[1.27\cdot 10^{9}\left(\frac{T_e}{10^4 \mbox{ K}}\right)^{0.59}
\left(\frac{\nu}{10^9 \mbox{ Hz}}\right)^{-0.1}
\frac{F_{\nu} (\mbox{Jy})}{F(\ha) (\mbox{erg s}^{-1}\mbox{ cm}^{-2})}\right] ,
\label{tau1}
\end{equation}

\noindent where $T_e$ is the electron temperature and $\nu$ is the frequency at
which the radio continuum flux, $F_{\nu}$, is measured. For NGC~4214, 
$T_e = 10\, 500$ K \citep{KobuSkil96}, and for $\nu$ = 4.885 GHz (this choice of
frequency will become obvious in the next paragraph) we obtain:

\begin{equation}
\tau_{\rm rad} = \ln\left[ \frac{F_{\nu} (\mbox{Jy})}
{F(\ha) (\mbox{erg s}^{-1}\mbox{ cm}^{-2})}\right] - 20.83 .
\label{tau2}
\end{equation}

	Equation~\ref{tau1} is not a strong function of temperature. For
example, changing $T_e$ by 1\, 000 K produces a change of only 0.05 or 0.06 in
$\tau_{\rm rad}$. 
The largest uncertainty by far when using this equation is discerning 
the contribution of non-thermal emission to the radio continuum. To do that, we
assume the presence of a thermal component with spectral index 
$\alpha_{\rm T} = 0.1$ and of a non-thermal one with spectral index 
$\alpha_{\rm NT}$ \citep{SramWeed86} and use our measured values of $\alpha$ 
to assign which fraction of the radio continuum corresponds to each 
one of them.  Unfortunately, the value of $\alpha_{\rm NT}$ 
can be as high as $\approx -0.6$ and as low as $\approx -1.4$. Therefore, 
only when $\alpha$ is close to $-0.1$ can $\tau_{\rm rad}$ be accurately
determined. 

	We show our values of $\tau_{\rm rad}$ measured from
Eq.~\ref{tau1} in Table~\ref{radioap} for three of the radio apertures. We
have not attempted to measure $\tau_{\rm rad}$ at the other two apertures due 
to dominance
of non-thermal emission there. Two values are given for I-A and II-BCD
considering the two possibilities for $\alpha_{\rm NT}$, while only one is given
for II-AE since the non-thermal contribution there is negligible. The measured
\ha\ flux used to calculate $\tau_{\rm rad}$ is also shown for those apertures 
in Table~\ref{radioap} and was obtained from the WFPC2 data after the resolution
was matched to that of the radio data. The \ha\ flux corrected for extinction 
using the Balmer ratio data of M99 and a foreground screen model is 
also shown, and from the logarithm of the ratio of the two quantities we obtain 
$\tau_{\rm rad}$ as measured using the Balmer ratio. 

	In the three cases we find that $\tau_{\rm rad}$ is significantly higher
than $\tau_{\rm Bal}$ (the apparent optical depth at \ha\ as measured from the 
Balmer ratio), which indicates that the uniform foreground screen model
is not valid in this case, as already found for other star-forming regions
(\citealt{Maizetal00a} and references therein). We will analyze the significance
of this result later on. We only point out here that the value of 
$\tau_{\rm rad}$ is the real mean optical depth at \ha\ and that it is 
independent of the resolution of the data as long as all the fluxes are 
included. On the other hand, the value of $\tau_{\rm Bal}$ can be heavily 
weighted toward low-extinction areas (holes in the foreground screen or skin 
emission from dusty \ion{H}{2} regions) and that its value is resolution 
dependent. 

\section{DISCUSSION}

	The WFPC2 nebular images of NGC 4214 presented here are an illuminating
example of the complex morphology of the ionized gas around regions with intense
star formation. This complexity is sometimes overlooked when more distant
unresolved objects are studied.

\subsection{Sources and sinks of ionizing photons}

	The balance between sources and sinks of ionizing photons in the ISM is 
not completely understood. Regarding the sources, several problems exist:
the estimation of the ionizing flux as a function of initial mass, the 
detection of the massive stars hidden in their dust enshrouded cocoons 
\citep{Drisetal00}, and the precise measurement of stellar masses around the 
high mass end of the IMF \citep{Antoetal00}. Of those problems, the most 
serious one is the first one: the extreme opacity of the ISM below 912~\AA\ 
hampers the direct measurement of the ionizing fluxes of O and B stars.
Even though a lot of effort has been made in the modelling of the ionizing
fluxes, the strong winds in the atmospheres of those stars and other
complications make the results somewhat uncertain (see, e.g., 
\citealt{SchadeKo97}). At the present time it has been possible to analyze the 
ionizing flux from only two B stars, $\beta$ and $\varepsilon$ CMa, with 
somewhat contradictory results \citep{Cassetal95,Cassetal96}, and no direct 
measurements exist for spectral type O. Therefore, it is not clear how many 
ionizing photons are emitted even in the case of well studied \ion{H}{2} 
regions. 

	Regarding the sinks of the ionizing photons three major questions remain
to be answered completely: What fraction of the ionizing photons is 
(a) destroyed by dust, (b) not absorbed locally but in the rest of the galaxy,
(c) escape from the galaxy? 
That dust plays a large role in the interaction of the ionizing flux with the
ISM is readily apparent in the HST images of M16 \citep{Hestetal96}. 
With respect to the fraction of photons that are absorbed elsewhere in the 
galaxy (to form the DIG), several studies place it close to 40\% 
independent of the Hubble type (\citealt{Greeetal98} and references 
therein). The number of photons which actually escape is more difficult to
measure and only a small sample has been studied. The results indicate that the
fraction is quite small \citep{Leitetal95}.

	Our images of NGC 4214 show that the number of ionizing photons escaping
from an \ion{H}{2} region can be highly variable. The solid angle covered by 
the surrounding dense gas around the SSC I-As is certainly not large and,
even though our three dimensional information is not complete, it is clear 
from the geometry that more than half of the ionizing photons from the SSC 
itself are not absorbed by the nearby surrounding gas but must escape to large 
distances, preferentially along directions close to the line of sight (towards
or away from us). Furthermore, the
low extinction observed in the UV \citep{MasHKunt99} and the likely puncture
of the galactic disk revealed by kinematic data (MMTM) suggest that a
significant fraction of the ionizing flux may be escaping to the intergalactic
medium. On the other hand, the dense clouds of gas which exist around the
ionizing clusters of II-A and II-B make it more difficult for the Lyman
continuum photons to escape far away and most of them should be contributing to 
the formation of Balmer lines close to their sources or be destroyed by dust.
With respect to the other massive clusters, I-Ds has even less gas around it
than I-As while II-C and I-Bs are in an intermediate situation between II-A or
II-B and I-As.

	Knots II-A and II-B are thus prototypes of massive compact \ion{H}{2} 
regions with core sizes of $\lesssim 20$ pc for the \ha\ emission. 
They are also distinguished by their high values of 
\oiiir\ / \ha\ and \wha, the high dust content of
their immediate surroundings, and the gaussian profile of their emission lines. 
As we have seen in the previous section, the high values of \wha\ indicate a 
small ($\lesssim 3$ Myr) age. These properties indicate that the gas is being 
ionized {\em in situ}, as we will discuss in more detail afterwards. Two other 
knots in NGC~4214 show very similar characteristics, the ones located in the 
small I-A1n and I-A3n apertures. They are clearly differentiated from their 
surroundings by their point-like character (as opposed to the filamentary 
structures around them), high intensity contrast, and excitation. They are also 
located next to dust clouds and the velocity profiles of their emission lines
is gaussian. The main difference between these knots and the ones in 
NGC 4214-II is that they are not located on top of the most massive cluster in 
the area (I-As), but at some distance from it (60 pc in projection for the knot 
in I-A1n and 120 pc for I-A3n). However, the similarities in the observed
properties indicate that the central knots in I-A1n and I-A3n are ionized 
{\em in situ} and not by the SSC. This is supported by the fact that they cover
relatively small solid angles as seen from I-As ($\approx$ 1.9\% and 0.5\%,
respectively, if they all are in the same plane perpendicular to the line of 
sight and are spherically symmetric), but contain four to eight times more of 
the fraction of the total \ha\ flux in NGC 4214 I-A ($8-11$\% and $2.5-4.0$\%,
respectively) and six to eleven times of the \oiiir\ flux ($12-19$\% and 
$3.0-5.5$\%, respectively). This situation is quite similar to the one found
in N11 in the LMC \citep{Parketal92,WalbPark92}, where an $\approx 4$ Myr old 
cluster has triggered the formation of a second cluster at the present 
\ha\ maximum. Those authors coined the term {\em two-stage starburst} to 
describe such an event, because the disruption of the ISM produced by the first 
cluster appears to be the cause for the formation of the second one. We propose 
here that I-As + (I-A1n + I-A3n) could be such an event, but certainty cannot be
reached unless UV spectroscopy of the individual knots is obtained. Other
examples of multi-stage starbursts in nearby galaxies are analyzed in 
\citet{GonDetal97} and \citet{Leit98} and references therein.

	We present in Table~\ref{binned} the \ha\ and \oiiir\ fluxes binned as
a function of pixel flux. We established the limit between the DIG and the 
\ion{H}{2} regions by averaging \ha\ and \oiiir\ and making intensity cuts as 
a function of position for different knots. We then determined that the change 
of slope at the border of the \ion{H}{2} regions takes place at
$\approx 0.3 \cdot 10^{-16}$ erg s$^{-1}$ cm$^{-2}$ for most cases. Thus, the 
sum of the first two bins corresponds to the percentage of the emission 
originating in the DIG and the value obtained (42\% for \ha\ and 37\% for 
\oiiir) is similar to that of other galaxies.

	The diffuse ionized gas is preferentially distributed in two areas
adjacent to the main \ion{H}{2} complexes: (a) in between the two complexes,
to the S of NGC 4214-I and W of NGC 4214-II (around I-Hn); and (b) at the N 
border of NGC 4214-I. The value of \oiiir\ / \ha\ in these two regions is quite 
different: in the first case we have already mentioned that high values are
found (the integrated value for the I-Hn aperture is close to 1.5, even higher
than the integrated value for NGC 4214-II), while the second area has lower 
excitation values, in the range (0.6-0.8) typical of most of the DIG in NGC 
4214. What is the origin of this difference? The most likely explanation is that
in the first area a population of dispersed OB stars is contributing most of the
ionizing photons, thus providing a nearby source, while in the second area 
the ionizing source is farther away (probably the stars in I-Ds as well as some
leakage form the rest of NGC 4214-I). The difference in distance to the 
source could cause the change in ionization parameter needed to explain the 
different excitation ratios. This possibility will be further explored in Paper
II.

\subsection{Bubbles and kinematic ages}

	A morphological evolution is readily apparent in our WFPC2 images of NGC
4214. The \ion{H}{2} regions around young clusters like II-A and II-B are 
compact and filled while the ones around older clusters like I-As and I-Bs are
much larger and have been evacuated in their central parts. An intermediate
case, where a small bubble can be observed around the cluster, is seen in
II-C while older clusters like I-Ds have little gas around them. This
evolutionary trend is consistent with the morphology observed in other nearby
massive \ion{H}{2} regions like \objectname{NGC 604} or \objectname{NGC 2363}.

	Qualitatively, this evolution is easily explained by the deposition of
kinetic energy into the surrounding ISM by massive stars in the form of winds
and supernova explosions. The quantitative predictions of the models have been
compared with observations of clusters with a few O stars in the LMC by
\citet{Oey96}, but relatively little effort has been made to verify them for
more massive (and generally more distant) clusters.
The extrapolation of the classic treatment of wind bubbles created by single
stars of \cite{Weavetal77} to massive clusters yields a surprising result: given
the weak dependence of the size of the bubble $R$ on the input kinetic energy
luminosity $L_k$ and the surrounding medium density $n_0$ 
($R\propto (L_k/n_0)^{0.2}t^{0.6}$), $R$ should be almost solely dependent on 
$t$ for a wide variety of circumstances. For 2-4 Myr old clusters with the 
number of massive stars typical of the ones described in this paper $R$ should 
be in the range of 50 to 125 pc. MMTM used ground-based data to show that the 
bubbles around I-As and I-Bs were smaller than expected while the ones around 
the clusters in NGC 4214-II could not be observed.

	With our higher resolution WFPC2 data we confirm that the sizes of the
bubbles around I-As and I-Bs are indeed smaller than expected by 
$\approx$ 30\% (corresponding to a kinematic age smaller by $\approx$ 50\%).
What is more surprising is the absence of an apparent bubble around II-A and
II-B (even with a pixel size of 2 pc) and the small size of the one detected 
around II-C ($R \approx 20$ pc). This discrepancy is in the same direction as
the one detected by \citet{Oey96} for smaller clusters in the LMC. The NGC~4214
results cannot be explained by changing the number of stars, the value of 
$n_0$, or the age of the cluster and, therefore, we must conclude that the 
model of \citet{Weavetal77} is not applicable in these cases.

	The explanation of this effect may lie in the pressurized character of 
the cores of the molecular clouds where massive clusters form. The external 
pressure slows down the expansion of the wind-blown cavities around individual 
stars and can restrict them to $\lesssim 1$ pc sizes during their first 1-2 
Myr. For isolated stars, this phase is known as the ultracompact \ion{H}{2} 
region stage and may be quite long lasting, since it is possible for them
to reach a stable state \citep{GarSFran96}. For compact clusters, however, the
situation must be different. The high number of bubbles created and the internal
motion of the stars must end up producing a superbubble which is rapidly 
thermalized by wind-wind interactions. This superbubble would finally be able 
to break out of the cocoon but only after $\approx 2$ Myr have elapsed since
the formation of the cluster. 

	This 2-3 Myr delay would nicely fit the observations
described in this paper as well as those made for other nearby massive clusters.
For example, R136 is located at the center of an incomplete R $\approx$ 10 pc
\ion{H}{2} shell which is the brightest \ha\ structure in 30 Doradus and
probably delimits the extension of the cavity created by the cluster. The
incomplete shell traces the photoevaporating surface of the surrounding
high density gas \citep{Scowetal98}, where a very young triggered second
generation is already found \citep{Walbetal99}. The age of R136 is close to 2 
Myr \citep{WalbBlad97,MassHunt98}, so the size of the superbubble predicted by 
the \citet{Weavetal77} model should be an order of magnitude larger. However, if
we include a 2 Myr delay the observations can be easily reproduced since a 10 
pc superbubble can be formed in a timescale of one or a few hundred thousand 
years.

	It appears then that the clusters in NGC 4214-II may be at a similar
stage to that of R136. However, the kinematics of the warm ionized gas in 30 
Doradus is extremely complicated with many kinematic components
\citep{ChuKenn94}, while that of NGC 4214-II is much simpler, with only
single component gaussians (MMTM). The reason for this difference is probably
twofold: (1) The 30 Doradus region appears to have had a complicated star
formation history in the past 10 Myr \citep{WalbBlad97}. Thus, the behavior of 
the warm ionized gas should be dominated not by the kinetic energy input from
R136 but from that of the previous star-forming episodes. (2) When examined in
detail, the high intensity \ha\ emitting areas may not be so different after
all. The high spatial and spectral resolution data of \citet{Melnetal99} 
reveal that the velocity profile of the 10 pc shell around R136 (the F in their 
Figure 1; note that the figure does not have an aspect ratio of unity) is
one of the only ones in 30 Doradus which is approximately gaussian. This is what
would be expected of a photoevaporating surface but not of a rapidly expanding
shell or shell fragment \citep{TenTetal96}. Therefore, an \ha\ velocity profile
of 30 Doradus stripped of the high velocity components energized by 
previous star-forming episodes may not be so different from that of NGC 
4214-II.

\subsection{Where is the dust?}

	The distribution of the dust in a massive \ion{H}{2} region 
with respect to the gas and the stars, and the final effect it has on the 
measured radiation field is not very clear. A simple star-like
extinction model does not work because of the extended nature of the line
emission and because the proximity between the scattering particles and the
source can remove photons from the observed beam but also bring them back in.
Some authors have proposed the use of the term attenuation in such a case
instead of extinction to differentiate the two phenomena. \citet{Faneetal88}
found that the UV continuum was not as strongly affected by dust as it is in
standard extinction laws for a given reddening measured from the Balmer lines.
\citet{Calzetal94} derived an attenuation law comparing optical and ultraviolet 
starburst spectra based on those results. The reason for the discrepancy is 
easily seen in a case like I-A in NGC 4214. Even
though the cause of the UV continuum and the Balmer lines is the same (the
existence of massive stars), their spatial distributions are very different.
Therefore, the amount of dust present around each of the sources and between
them and us can also be very different. If the dust were located far away from
the object, the variations would be random; in some cases there would be
more dust present in front of the UV continuum sources than in front of the
ionized gas, and in other cases the opposite would be true. However, if the dust
were associated with the object, since the massive stars can evaporate (with 
their photons) and sweep (with their winds) the dust grains, there would be 
less dust in front of the stars than in front of the gas. Furthermore, a second
generation of stars can be produced in the swept-up gas $2-3$ Myr after the 
formation of the main $3-5$ Myr old burst \citep{Parketal92,WalbPark92}. The 
ionizing photons from those younger stars can provide a significant 
contribution to the total number of optical nebular photons emitted (since 
there is plenty of gas around them and early O stars have not yet disappeared) 
while being very hard to detect in the UV.

	The processes described in the previous paragraph are the physical
explanation of the \citet{Calzetal94} law. However, as we have seen, the 
younger massive \ion{H}{2} regions may not have had enough time to create a
superbubble around the cluster and in that case the \citet{Calzetal94} law
should not be applicable.

	\citet{Maizetal00a} give the expected values of $\tau_{\rm Bal}$ and
$\tau_{\rm rad}$ for three dust-gas configurations: uniform screen, patchy
screen, and dust-gas mixed. Our combination of radio and optical data shows that
a uniform dust screen model can not reproduce the observed values. A model with
dust and gas uniformly mixed gives a better fit but, in general, 
$\tau_{\rm rad}$ is too large for a given $\tau_{\rm Bal}$, specially for the
case of II-BCD. This is not surprising, given the existence of large structural 
non-uniformities inside the apertures used, as evidenced by the WFPC2 images. 
For the II-BCD aperture, the long-slit data of MMMVC indicate that we are 
including two regions (II-BD vs. II-C) with different attenuations, so the 
departure from the uniformly mixed model is larger than for II-AE or I-A, where
most of the Balmer emission is coming from a more homogeneous source. This is
especially true for II-AE, where the uniformly mixed model gives a reasonably 
good fit. Therefore, we can conclude that a significant fraction of the observed
dust (and maybe most of it) is physically associated with the ionized gas and is
not located far away from it. This result agrees with the detection in NGC 4214 
by MMMVC of small localized sources of {\em bluened} (i.e. with an \ha\ / \hb\
ratio lower than 2.86) Balmer radiation close to highly reddened sources. 

	This result agrees with the values observed in different parts of
\objectname{NGC 604} by \citet{Maizetal00a} and with the ones obtained by 
\citet{CaplDeha86} for a number of \ion{H}{2} regions in the \objectname{LMC} 
in the sense that a uniform screen cannot explain the measured optical and 
radio properties. The real optical depths at \ha\ appear to be larger by 0.5 to 
1.0 than the ones measured from the Balmer ratio. In the case of 
\objectname{NGC 5253}, the difference appears to be even larger
\citep{Calzetal97}. Therefore, a measurement of the number of ionizing photons 
obtained by measuring \ha\ and correcting for extinction with \hb\ may be
underestimated by a factor of $\approx 2$ or even more. This source of 
uncertainty should be taken into account when studying unresolved galaxies 
located much farther away than NGC 4214.

\section{CONCLUSIONS}

	We have analyzed WFPC2 and VLA images of NGC 4214. The main results we
have obtained can be summarized as follows:

\begin{itemize}
  \item A morphological evolution is observed in the star-forming knots of NGC
	4214 from filled compact \ion{H}{2} regions to shell-like objects to
	low intensity diffuse knots.
  \item Parallel to the morphological one, a similar evolution in the excitation
	ratios is detected as a function of age.
  \item A SNR has been discovered from the radio and optical data.
  \item A two-stage starburst candidate has been detected.
  \item The DIG in NGC 4214 contributes $\approx$ 40\% of the total 
	\ha\ and \oiiir\ emission of the galaxy.
  \item The pressurized character of the molecular cloud cores apparently 
	inhibits the formation of superbubbles around massive clusters for
	$\approx$ 2 Myr.
  \item Most of the dust detected in NGC 4214 by light attenuation around 
	massive clusters is physically associated with the ionized gas.
  \item The number of ionizing photons obtained from integrated optical 
 	spectra of galaxies similar to NGC 4214 may be underestimated by a
	factor of two due to the invalidity of the foreground screen 
	approximation for the reddening correction.
\end{itemize}

\acknowledgments

The authors would like to thank Miguel Cervi\~no for his help with the 
evolutionary synthesis models, Antx\'on Alberdi and Greg Taylor for their help 
with the VLA data analysis and J. Miguel Mas-Hesse for his useful insights on 
NGC 4214. We would also like to thank Richard Burg, Rosemary F.G. Wyse, and 
Richard E. Griffiths for encouragement and stimulating conversations. Finally,
we would like to thank Claus Leitherer and an anonymous referee for the useful
comments which helped improve this paper. Support for this 
work was provided by NASA through grant GO-06569.01-95A from the 
Space Telescope Science Institute, Inc., under NASA contract NAS5-26555.


\bibliographystyle{aj}
\bibliography{general}

\singlespace

\addtolength{\topmargin}{-0.35in}
\addtolength{\topmargin}{0.35in}

\include{mackenty.tab1}
\addtolength{\oddsidemargin}{-0.3in}
\addtolength{\evensidemargin}{-0.3in}
\include{mackenty.tab2}
\addtolength{\oddsidemargin}{-0.2in}
\addtolength{\evensidemargin}{-0.2in}
\addtolength{\topmargin}{-0.35in}
\include{mackenty.tab3}
\addtolength{\topmargin}{-0.35in}
\addtolength{\oddsidemargin}{+0.5in}
\addtolength{\evensidemargin}{+0.5in}
\include{mackenty.tab4}
\addtolength{\topmargin}{+0.7in}
\include{mackenty.tab5}
\include{mackenty.tab6}
\include{mackenty.tab7}

\end{document}

%% file: mackenty.tab1.tex
\begin{deluxetable}{llll}
\tablecaption{Cycle 6 HST Observations of NGC 4214.\label{wfpc2obs}}
\tabletypesize{\small}
\tablewidth{0pt}
\tablehead{\colhead{Filter}    & \colhead{Band} & Images & Exp. time (s)}
\startdata
F656N & \ha          & u3n8010fm + gm      & 800 + 800       \\
F502N & \oiiir       & u3n8010dm + em      & 700 + 800       \\
F336W & WFPC2 U      & u3n80101m + 2m + 3m & 260 + 900 + 900 \\
F555W & WFPC2 V      & u3n80104m + 5m + 6m & 100 + 600 + 600 \\
F702W & WFPC2 wide R & u3n80107m + 8m      & 500 + 500       \\
F814W & WFPC2 I      & u3n8010am + bm + cm & 100 + 600 + 600 \\
\enddata
\end{deluxetable}

%% file: mackenty.tab2.tex
\begin{deluxetable}{llcrrrcc}
\tablecaption{Unit nomenclature and corresponding circular apertures used in this work.\label{apertures}}
\tabletypesize{\scriptsize}
\tablewidth{0pt}
\tablehead{\colhead{name} & \colhead{old name/description\tablenotemark{a}} & \colhead{chip\tablenotemark{b}} & \colhead{$x$\tablenotemark{c}} & \colhead{$y$\tablenotemark{c}} & \colhead{radius} & \colhead{$\alpha$ (J2000)} & \colhead{$\delta$ (J2000)} \\
 & & & \colhead{(pix.)} & \colhead{(pix.)} & \colhead{(\arcsec)} & \colhead{12$^{\rm h}$15$^{\rm m}$+} & \colhead{36\arcdeg +} }
\startdata
I-A1n  & knot 3 (small aperture)                & 3 &  662.50 &  880.50 &  0.87 & 39\fs 214 & 19\arcmin 33\farcs 75      \\*
I-A1n  & knot 3 (interm. aperture)              & 3 &  668.25 &  879.75 &  1.94 & 39\fs 250 & 19\arcmin 33\farcs 39      \\*
I-A2n  & knot 10                                & 3 &  628.25 &  816.25 &  3.83 & 39\fs 397 & 19\arcmin 40\farcs 60      \\*
I-A3n  & knot 8 (small aperture)                & 3 &  618.00 &  877.00 &  0.87 & 38\fs 983 & 19\arcmin 37\farcs 17      \\*
I-A3n  & knot 8 (interm. aperture)              & 3 &  624.75 &  870.25 &  2.64 & 39\fs 060 & 19\arcmin 37\farcs 14      \\*
I-A4n  & knot 9                                 & 3 &  677.25 &  812.25 &  2.84 & 39\fs 697 & 19\arcmin 37\farcs 38      \\*
I-A5n  & knot 7                                 & 3 &  703.25 &  860.75 &  2.09 & 39\fs 558 & 19\arcmin 32\farcs 20      \\*
I-As   & knot A (small aperture)                & 3 &  672.75 &  851.75 &  1.34 & 39\fs 441 & 19\arcmin 34\farcs 99      \\*
I-A    & knot A (large aperture)                & 3 &  662.25 &  839.25 &  8.12 & 39\fs 456 & 19\arcmin 36\farcs 59 \\[1 mm]
I-B1n  & knot 5                                 & 2 &  768.75 &  819.25 &  2.49 & 40\fs 157 & 19\arcmin 30\farcs 18      \\*
I-B2n  & knot 6                                 & 2 &  800.75 &  785.25 &  2.24 & 40\fs 539 & 19\arcmin 30\farcs 25      \\*
I-Bs   & knot B (small aperture)                & 2 &  774.75 &  783.75 &  2.29 & 40\fs 400 & 19\arcmin 32\farcs 18      \\*
I-B    & knot B (large aperture)                & 2 &  794.75 &  794.75 &  6.32 & 40\fs 449 & 19\arcmin 30\farcs 03 \\[1 mm]
I-C1n  & knot W of knot 3*                      & 3 &  645.00 &  944.00 &  2.49 & 38\fs 741 & 19\arcmin 30\farcs 63      \\*
I-Cn   & W region of NW~complex*                & 3 &  649.00 &  976.00 &  6.47 & 38\fs 577 & 19\arcmin 28\farcs 15 \\[1 mm]
I-D1n  & knot 12 (N)                            & 3 &  543.75 &  783.25 &  1.44 & 39\fs 115 & 19\arcmin 48\farcs 91      \\*
I-D2n  & knot 12 (S)                            & 3 &  568.75 &  787.25 &  1.29 & 39\fs 232 & 19\arcmin 46\farcs 85      \\*
I-Ds   & knot C                                 & 3 &  522.25 &  767.75 &  1.74 & 39\fs 086 & 19\arcmin 51\farcs 51      \\*
I-D    & knot C-12                              & 3 &  543.75 &  762.25 &  6.37 & 39\fs 240 & 19\arcmin 50\farcs 34 \\[1 mm]
I-Es   & knot F                                 & 3 &  712.75 &  767.25 &  2.29 & 40\fs 155 & 19\arcmin 37\farcs 93      \\*
I-F    & knot I                                 & 3 &  718.25 &  801.25 &  2.29 & 39\fs 990 & 19\arcmin 35\farcs 21      \\*
I-Gn   & knot 11                                & 2 &  851.25 &  741.75 &  1.69 & 41\fs 083 & 19\arcmin 29\farcs 66      \\*
I-Hn   & froth                                  & 2 &  783.50 &  924.00 &  1.99 & 39\fs 624 & 19\arcmin 21\farcs 98 \\[1 mm]
I      & NW complex                             & 3 &  700.00 &  836.00 & 20.62 & 39\fs 686 & 19\arcmin 34\farcs 12 \\[1 mm]
II-A   & knot 2-K                               & 2 & 1041.75 &  924.75 &  3.14 & 41\fs 097 & 19\arcmin 03\farcs 60      \\*
II-B   & knot 1-D                               & 2 &  963.75 &  919.25 &  1.69 & 40\fs 683 & 19\arcmin 09\farcs 54      \\*
II-C1n & knot 4-E (S)                           & 2 &  930.00 &  907.75 &  1.27 & 40\fs 557 & 19\arcmin 12\farcs 73      \\*
II-C2n & knot 4-E (N)                           & 2 &  927.25 &  888.75 &  1.15 & 40\fs 654 & 19\arcmin 14\farcs 23      \\*
II-C   & knot 4-E                               & 2 &  926.25 &  898.25 &  2.59 & 40\fs 592 & 19\arcmin 13\farcs 65      \\*
II-Dn  & knot L (N)                             & 2 &  982.75 &  938.75 &  1.49 & 40\fs 676 & 19\arcmin 06\farcs 84      \\*
II-En  & knot L (S)                             & 2 & 1009.25 &  956.25 &  1.59 & 40\fs 725 & 19\arcmin 03\farcs 74 \\[1 mm]
II     & SE complex                             & 2 &  977.75 &  937.25 &  9.96 & 40\fs 657 & 19\arcmin 07\farcs 30 \\[1 mm]
IIIs   & nucleus*                               & 3 &  496.50 &  897.75 &  1.44 & 38\fs 174 & 19\arcmin 44\farcs 44 \\[1 mm]
IVs    & old cluster?*                          & 3 &  327.50 &  896.25 &  1.47 & 37\fs 222 & 19\arcmin 56\farcs 66 \\[1 mm]
V-An   & Far W region (high exc. zone)*         & 3 &  111.00 & 1318.00 &  4.78 & 33\fs 494 & 19\arcmin 43\farcs 10      \\*
Vn     & Far W region*                          & 3 &  183.00 & 1203.00 & 17.53 & 34\fs 578 & 19\arcmin 45\farcs 88      \\*
VI-An  & Far S region (high exc. zone)*         & 2 &  964.00 & 1199.00 &  4.08 & 39\fs 031 & 18\arcmin 50\farcs 14      \\*
VIn    & Far S region*                          & 2 &  895.00 & 1267.00 & 14.14 & 38\fs 236 & 18\arcmin 50\farcs 35      \\*
VIIn   & knot 15                                & 2 & 1015.00 &  813.00 &  5.48 & 41\fs 602 & 19\arcmin 13\farcs 20      \\*
VIIIn  & knot 14                                & 2 &  917.00 &  808.00 &  4.98 & 41\fs 070 & 19\arcmin 20\farcs 50      \\*
IXn    & knot 13                                & 2 & 1150.00 &  824.00 &  5.58 & 42\fs 309 & 19\arcmin 02\farcs 83      \\*
Xn     & old shell*                             & 2 & 1398.00 &  957.00 &  7.97 & 42\fs 938 & 18\arcmin 35\farcs 98      \\*
XIn    & knot E of old cluster*                 & 3 &  200.00 &  915.00 &  4.98 & 36\fs 383 & 20\arcmin 04\farcs 47      \\*
XIIn   & high exc. region N of knot 12          & 4 &  380.00 &  723.00 &  3.29 & 38\fs 540 & 20\arcmin 04\farcs 62      \\*
XIIIn  & high exc. region E of knot 15*         & 2 & 1040.00 &  745.00 &  2.39 & 42\fs 143 & 19\arcmin 16\farcs 09         
\enddata
\tablenotetext{a}{See \citet{Maizetal98} or \citet{Maiz99}.}
\tablenotetext{b}{WFPC2 chip at the aperture center.}
\tablenotetext{c}{Coordinates of the aperture center in our mosaics.}
\tablenotetext{*}{Extinction uncertain.}
\end{deluxetable}

%% file: mackenty.tab3.tex
\begin{deluxetable}{lrrrrrrrrrrrrrrrr}
\tablecaption{Integrated \ha, \oiiir, and \ha\ continuum fluxes obtained from different extinction models for the apertures in Table~\ref{apertures}.\label{intfluxes}}
\tabletypesize{\scriptsize}
\tablewidth{0pt}
\tablehead{\colhead{name} & & \multicolumn{5}{c}{log $F$(\ha)} & & \multicolumn{5}{c}{log $F$([O III] $\lambda$5007)} & & \multicolumn{3}{c}{log $F$($\lambda$6563)} \\
 & & \multicolumn{5}{c}{(ergs s$^{-1}$ cm$^{-2}$)} & & \multicolumn{5}{c}{(ergs s$^{-1}$ cm$^{-2}$)} & & \multicolumn{3}{c}{(ergs s$^{-1}$ cm$^{-2}$ \AA$^{-1}$)} \\
 & \phm{0} & \colhead{no ext.\tablenotemark{a}} & \colhead{ext. 1\tablenotemark{b}} & \colhead{ext. 2\tablenotemark{c}} & \colhead{\% 1\tablenotemark{d}} & \colhead{\% 2\tablenotemark{e}} & \phm{0} & \colhead{no ext.\tablenotemark{a}} & \colhead{ext. 1\tablenotemark{b}} & \colhead{ext. 2\tablenotemark{c}} & \colhead{\% 1\tablenotemark{d}} & \colhead{\% 2\tablenotemark{e}} & \phm{0} & \colhead{no ext.\tablenotemark{a}} & \colhead{ext. 1\tablenotemark{b}} & \colhead{\% 1\tablenotemark{d}} }
\startdata
I-A1n  &  & -12.68 & -12.42 & -12.29 &   84 &  151 &  & -12.60 & -12.23 & -12.05 &  137 &  254 &  & -15.82 & -15.55 &   86      \\*
I-A1n  &  & -12.30 & -12.11 & -12.02 &   54 &   92 &  & -12.27 & -11.99 & -11.86 &   90 &  159 &  & -15.20 & -15.05 &   42      \\*
I-A2n  &  & -12.40 & -12.26 & -12.22 &   37 &   48 &  & -12.56 & -12.37 & -12.32 &   54 &   73 &  & -14.71 & -14.55 &   44      \\*
I-A3n  &  & -13.17 & -12.86 & -12.72 &  103 &  183 &  & -13.20 & -12.77 & -12.59 &  165 &  302 &  & -16.13 & -15.81 &  106      \\*
I-A3n  &  & -12.47 & -12.25 & -12.15 &   67 &  110 &  & -12.58 & -12.27 & -12.13 &  106 &  182 &  & -15.05 & -14.86 &   56      \\*
I-A4n  &  & -12.54 & -12.35 & -12.24 &   53 &  100 &  & -12.68 & -12.43 & -12.29 &   78 &  148 &  & -15.06 & -14.87 &   54      \\*
I-A5n  &  & -12.56 & -12.46 & -12.44 &   28 &   32 &  & -12.62 & -12.47 & -12.45 &   38 &   47 &  & -15.13 & -15.01 &   32      \\*
I-As   &  & -13.22 & -13.17 & -13.16 &   12 &   13 &  & -13.22 & -13.15 & -13.14 &   17 &   20 &  & -14.84 & -14.80 &   10      \\*
I-A    &  & -11.58 & -11.40 & -11.32 &   50 &   79 &  & -11.68 & -11.43 & -11.32 &   77 &  130 &  & -14.05 & -13.89 &   43 \\[1 mm]
I-B1n  &  & -12.39 & -12.28 & -12.27 &   27 &   32 &  & -12.51 & -12.37 & -12.34 &   39 &   49 &  & -15.12 & -15.02 &   24      \\*
I-B2n  &  & -12.52 & -12.34 & -12.29 &   52 &   69 &  & -12.62 & -12.38 & -12.31 &   75 &  104 &  & -15.14 & -14.99 &   41      \\*
I-Bs   &  & -12.53 & -12.41 & -12.39 &   31 &   37 &  & -12.63 & -12.46 & -12.42 &   47 &   59 &  & -14.98 & -14.86 &   30      \\*
I-B    &  & -11.81 & -11.68 & -11.61 &   37 &   61 &  & -11.95 & -11.76 & -11.68 &   52 &   85 &  & -14.39 & -14.27 &   33 \\[1 mm]
I-C1n  &  & -12.79 & -12.74 & -12.73 &   12 &   13 &  & -12.74 & -12.67 & -12.66 &   17 &   19 &  & -15.26 & -15.21 &   12      \\*
I-Cn   &  & -12.29 & -12.23 & -12.21 &   14 &   21 &  & -12.37 & -12.28 & -12.25 &   21 &   30 &  & -14.54 & -14.48 &   15 \\[1 mm]
I-D1n  &  & -13.11 & -13.00 & -12.93 &   30 &   53 &  & -13.14 & -12.99 & -12.92 &   41 &   67 &  & -15.79 & -15.65 &   37      \\*
I-D2n  &  & -13.27 & -13.12 & -12.91 &   41 &  128 &  & -13.32 & -13.12 & -12.89 &   61 &  169 &  & -15.90 & -15.72 &   49      \\*
I-Ds   &  & -13.45 & -13.30 & -13.28 &   39 &   49 &  & -13.56 & -13.38 & -13.33 &   53 &   69 &  & -15.38 & -15.29 &   25      \\*
I-D    &  & -12.37 & -12.22 & -12.13 &   41 &   74 &  & -12.53 & -12.33 & -12.21 &   59 &  107 &  & -14.56 & -14.40 &   43 \\[1 mm]
I-Es   &  & -12.87 & -12.72 & -12.69 &   41 &   51 &  & -13.20 & -12.99 & -12.94 &   61 &   80 &  & -15.29 & -15.11 &   49      \\*
I-F    &  & -12.56 & -12.43 & -12.41 &   35 &   41 &  & -12.81 & -12.63 & -12.59 &   51 &   64 &  & -15.20 & -15.07 &   36      \\*
I-Gn   &  & -13.32 & -12.95 & -11.90 &  132 & 2543 &  & -13.36 & -12.86 & -11.72 &  214 & 4177 &  & -15.85 & -15.50 &  123      \\*
I-Hn   &  & -13.43 & -13.31 & -13.29 &   31 &   36 &  & -13.30 & -13.14 & -13.10 &   46 &   58 &  & -15.67 & -15.56 &   29 \\[1 mm]
I      &  & -11.17 & -11.02 & -10.89 &   40 &   88 &  & -11.29 & -11.09 & -10.90 &   60 &  144 &  & -13.48 & -13.36 &   33 \\[1 mm]
II-A   &  & -12.23 & -11.97 & -11.71 &   83 &  232 &  & -12.17 & -11.82 & -11.52 &  124 &  349 &  & -15.19 & -14.93 &   83      \\*
II-B   &  & -12.39 & -12.26 & -12.10 &   32 &   93 &  & -12.35 & -12.16 & -11.89 &   53 &  186 &  & -15.52 & -15.40 &   33      \\*
II-C1n &  & -12.75 & -12.72 & -12.72 &    8 &    9 &  & -12.76 & -12.72 & -12.71 &   10 &   12 &  & -15.69 & -15.66 &    7      \\*
II-C2n &  & -13.03 & -13.01 & -13.01 &    4 &    4 &  & -13.06 & -13.04 & -13.03 &    5 &    5 &  & -15.91 & -15.89 &    3      \\*
II-C   &  & -12.41 & -12.35 & -12.33 &   14 &   20 &  & -12.46 & -12.39 & -12.36 &   18 &   26 &  & -15.26 & -15.19 &   18      \\*
II-Dn  &  & -12.68 & -12.55 & -12.53 &   34 &   41 &  & -12.63 & -12.45 & -12.42 &   49 &   61 &  & -15.75 & -15.62 &   36      \\*
II-En  &  & -12.69 & -12.55 & -12.53 &   37 &   44 &  & -12.66 & -12.47 & -12.44 &   54 &   67 &  & -15.74 & -15.60 &   39 \\[1 mm]
II     &  & -11.59 & -11.43 & -11.24 &   43 &  121 &  & -11.59 & -11.36 & -11.09 &   69 &  216 &  & -14.28 & -14.15 &   37 \\[1 mm]
IIIs   &  & -13.85 & -13.80 & -13.80 &   12 &   13 &  & -13.99 & -13.92 & -13.91 &   16 &   19 &  & -15.05 & -15.00 &   12 \\[1 mm]
IVs    &  & -15.28 & -15.23 & -15.22 &   12 &   13 &  & -14.59 & -14.52 & -14.51 &   17 &   19 &  & -15.30 & -15.25 &   12 \\[1 mm]
V-An   &  & -13.41 & -13.36 & -13.36 &   12 &   13 &  & -13.23 & -13.16 & -13.15 &   17 &   19 &  & -15.90 & -15.85 &   12      \\*
Vn     &  & -12.25 & -12.20 & -12.19 &   12 &   13 &  & -12.47 & -12.40 & -12.39 &   17 &   19 &  & -14.24 & -14.19 &   12      \\*
VI-An  &  & -13.46 & -13.41 & -13.40 &   12 &   13 &  & -13.35 & -13.28 & -13.27 &   17 &   19 &  & -15.58 & -15.53 &   12      \\*
VIn    &  & -12.53 & -12.48 & -12.48 &   12 &   13 &  & -12.57 & -12.50 & -12.49 &   17 &   19 &  & -14.54 & -14.49 &   12      \\*
VIIn   &  & -12.65 & -12.52 & -12.40 &   35 &   80 &  & -12.95 & -12.77 & -12.63 &   51 &  112 &  & -14.87 & -14.75 &   33      \\*
VIIIn  &  & -12.73 & -12.65 & -12.63 &   19 &   24 &  & -12.80 & -12.70 & -12.67 &   27 &   35 &  & -14.83 & -14.74 &   22      \\*
IXn    &  & -12.78 & -12.59 & -12.31 &   55 &  196 &  & -12.87 & -12.60 & -12.25 &   83 &  309 &  & -15.20 & -14.99 &   62      \\*
Xn     &  & -13.03 & -12.98 & -12.98 &   12 &   13 &  & -13.34 & -13.27 & -13.26 &   17 &   19 &  & -15.09 & -15.04 &   12      \\*
XIn    &  & -13.40 & -13.35 & -13.35 &   12 &   13 &  & -13.72 & -13.65 & -13.64 &   16 &   19 &  & -15.06 & -15.00 &   12      \\*
XIIn   &  & -13.45 & -13.39 & -13.39 &   13 &   14 &  & -13.46 & -13.37 & -13.36 &   20 &   25 &  & -15.32 & -15.27 &   12      \\*
XIIIn  &  & -13.67 & -13.44 & -13.24 &   72 &  173 &  & -13.48 & -13.12 & -12.87 &  127 &  309 &  & -15.99 & -15.75 &   71 \\[1 mm]
All &  & -10.86 & -10.74 & -10.62 &   33 &   76 &  & -10.96 & -10.78 & -10.59 &   52 &  133 &  & -12.90 & -12.82 &   20
\enddata
\tablenotetext{a}{No extinction correction applied.}
\tablenotetext{b}{Foreground screen correction applied.}
\tablenotetext{c}{Gas/dust mixed correction applied.}
\tablenotetext{d}{Percentage increase between first and second columns.}
\tablenotetext{e}{Percentage increase between first and third columns.}
\end{deluxetable}

%% file: mackenty.tab4.tex
\begin{deluxetable}{lrrrrrrrrrrrrrrrr}
\tablecaption{Integrated \oiiir/\ha\ and \wha\ obtained from different extinction models for the apertures in Table~\ref{apertures}.\label{intratios}}
\tabletypesize{\scriptsize}
\tablewidth{0pt}
\tablehead{\colhead{name} & \multicolumn{7}{c}{[O III] $\lambda$5007 /\ha} & \multicolumn{7}{c}{$W$(\ha) (\AA)} \\
 & \phm{0} & \colhead{no} & \colhead{ext.} & \colhead{ext.} & \colhead{\%} & \colhead{\%} & \phm{0} & \colhead{no} & \colhead{line} & \colhead{line} & \colhead{all} & \colhead{\% line} & \colhead{\% line} & \colhead{\% all} \\
 & & \colhead{ext.\tablenotemark{a}} & \colhead{1\tablenotemark{b}} & \colhead{2\tablenotemark{c}} & \colhead{1\tablenotemark{d}} & \colhead{2\tablenotemark{e}} & & \colhead{ext.\tablenotemark{a}} & \colhead{ext. 1\tablenotemark{b}} & \colhead{ext. 2\tablenotemark{c}} & \colhead{ext.\tablenotemark{f}} & \colhead{ext. 1\tablenotemark{d}} & \colhead{ext. 2\tablenotemark{e}} & \colhead{ext.\tablenotemark{g}} }
\startdata
I-A1n  &  & 1.21 & 1.56 & 1.71 &   28 &   41 &  & 1368 & 2521 & 3433 & 1348 &   84 &  151 &   -1      \\*
I-A1n  &  & 1.07 & 1.32 & 1.45 &   22 &   34 &  &  791 & 1222 & 1519 &  858 &   54 &   92 &    8      \\*
I-A2n  &  & 0.68 & 0.76 & 0.80 &   12 &   17 &  &  206 &  283 &  305 &  196 &   37 &   48 &   -4      \\*
I-A3n  &  & 0.95 & 1.23 & 1.34 &   30 &   41 &  &  900 & 1831 & 2554 &  888 &  103 &  183 &   -1      \\*
I-A3n  &  & 0.77 & 0.94 & 1.03 &   23 &   34 &  &  381 &  638 &  801 &  408 &   67 &  110 &    7      \\*
I-A4n  &  & 0.72 & 0.84 & 0.89 &   16 &   24 &  &  328 &  504 &  656 &  327 &   53 &  100 &    0      \\*
I-A5n  &  & 0.88 & 0.96 & 0.98 &    8 &   11 &  &  372 &  477 &  493 &  360 &   28 &   32 &   -3      \\*
I-As   &  & 0.99 & 1.03 & 1.05 &    4 &    6 &  &   42 &   48 &   48 &   43 &   12 &   13 &    1      \\*
I-A    &  & 0.80 & 0.94 & 1.02 &   18 &   27 &  &  294 &  442 &  529 &  308 &   50 &   79 &    4 \\[1 mm]
I-B1n  &  & 0.75 & 0.82 & 0.84 &    9 &   12 &  &  535 &  683 &  707 &  551 &   27 &   32 &    2      \\*
I-B2n  &  & 0.79 & 0.91 & 0.95 &   15 &   20 &  &  422 &  642 &  716 &  454 &   52 &   69 &    7      \\*
I-Bs   &  & 0.80 & 0.89 & 0.92 &   11 &   15 &  &  282 &  371 &  388 &  286 &   31 &   37 &    1      \\*
I-B    &  & 0.74 & 0.82 & 0.84 &   11 &   14 &  &  378 &  519 &  611 &  389 &   37 &   61 &    3 \\[1 mm]
I-C1n  &  & 1.12 & 1.17 & 1.18 &    4 &    5 &  &  300 &  336 &  339 &  299 &   12 &   13 &    0      \\*
I-Cn   &  & 0.84 & 0.89 & 0.91 &    5 &    7 &  &  176 &  202 &  212 &  175 &   14 &   21 &    0 \\[1 mm]
I-D1n  &  & 0.93 & 1.02 & 1.02 &    8 &    9 &  &  477 &  620 &  730 &  451 &   30 &   53 &   -5      \\*
I-D2n  &  & 0.89 & 1.01 & 1.05 &   13 &   18 &  &  424 &  601 &  967 &  402 &   41 &  128 &   -5      \\*
I-Ds   &  & 0.77 & 0.85 & 0.88 &   10 &   13 &  &   86 &  120 &  128 &   96 &   39 &   49 &   11      \\*
I-D    &  & 0.69 & 0.78 & 0.83 &   12 &   18 &  &  154 &  219 &  270 &  153 &   41 &   74 &    0 \\[1 mm]
I-Es   &  & 0.47 & 0.53 & 0.56 &   14 &   19 &  &  261 &  369 &  394 &  247 &   41 &   51 &   -5      \\*
I-F    &  & 0.56 & 0.63 & 0.65 &   12 &   16 &  &  440 &  595 &  622 &  434 &   35 &   41 &   -1      \\*
I-Gn   &  & 0.92 & 1.25 & 1.49 &   35 &   61 &  &  340 &  791 & 8999 &  353 &  132 & 2543 &    3      \\*
I-Hn   &  & 1.33 & 1.49 & 1.54 &   12 &   16 &  &  175 &  229 &  238 &  176 &   31 &   36 &    1 \\[1 mm]
I      &  & 0.76 & 0.87 & 0.98 &   14 &   29 &  &  206 &  289 &  388 &  216 &   40 &   88 &    5 \\[1 mm]
II-A   &  & 1.16 & 1.41 & 1.57 &   21 &   35 &  &  902 & 1660 & 2999 &  907 &   83 &  232 &    0      \\*
II-B   &  & 1.08 & 1.26 & 1.60 &   16 &   48 &  & 1368 & 1807 & 2647 & 1353 &   32 &   93 &   -1      \\*
II-C1n &  & 0.99 & 1.01 & 1.02 &    1 &    2 &  &  871 &  943 &  953 &  876 &    8 &    9 &    0      \\*
II-C2n &  & 0.93 & 0.94 & 0.94 &    0 &    1 &  &  753 &  786 &  789 &  756 &    4 &    4 &    0      \\*
II-C   &  & 0.88 & 0.91 & 0.93 &    3 &    5 &  &  713 &  818 &  857 &  690 &   14 &   20 &   -3      \\*
II-Dn  &  & 1.14 & 1.26 & 1.30 &   10 &   14 &  & 1179 & 1589 & 1664 & 1164 &   34 &   41 &   -1      \\*
II-En  &  & 1.07 & 1.20 & 1.24 &   12 &   16 &  & 1117 & 1541 & 1613 & 1108 &   37 &   44 &    0 \\[1 mm]
II     &  & 0.98 & 1.16 & 1.41 &   17 &   43 &  &  498 &  715 & 1100 &  521 &   43 &  121 &    4 \\[1 mm]
IIIs   &  & 0.73 & 0.76 & 0.77 &    3 &    5 &  &   16 &   18 &   18 &   16 &   12 &   13 &    0 \\[1 mm]
IVs    &  & 4.88 & 5.09 & 5.15 &    4 &    5 &  &    1 &    1 &    1 &    1 &   12 &   13 &    0 \\[1 mm]
V-An   &  & 1.53 & 1.59 & 1.61 &    4 &    5 &  &  304 &  341 &  344 &  304 &   12 &   13 &    0      \\*
Vn     &  & 0.60 & 0.62 & 0.63 &    4 &    5 &  &   99 &  111 &  112 &   99 &   12 &   13 &    0      \\*
VI-An  &  & 1.28 & 1.34 & 1.36 &    4 &    5 &  &  133 &  149 &  150 &  133 &   12 &   13 &    0      \\*
VIn    &  & 0.91 & 0.95 & 0.96 &    4 &    5 &  &  103 &  116 &  117 &  103 &   12 &   13 &    0      \\*
VIIn   &  & 0.50 & 0.56 & 0.59 &   11 &   17 &  &  166 &  226 &  299 &  169 &   35 &   80 &    1      \\*
VIIIn  &  & 0.84 & 0.89 & 0.91 &    6 &    8 &  &  127 &  151 &  158 &  124 &   19 &   24 &   -2      \\*
IXn    &  & 0.82 & 0.97 & 1.13 &   18 &   38 &  &  265 &  411 &  785 &  252 &   55 &  196 &   -4      \\*
Xn     &  & 0.49 & 0.51 & 0.52 &    4 &    5 &  &  116 &  130 &  131 &  116 &   12 &   13 &    0      \\*
XIn    &  & 0.48 & 0.50 & 0.51 &    4 &    5 &  &   45 &   51 &   51 &   45 &   12 &   13 &    0      \\*
XIIn   &  & 0.98 & 1.04 & 1.07 &    7 &    9 &  &   75 &   84 &   86 &   75 &   13 &   14 &    0      \\*
XIIIn  &  & 1.55 & 2.05 & 2.33 &   31 &   49 &  &  205 &  355 &  562 &  206 &   72 &  173 &    0 \\[1 mm]
All &  & 0.80 & 0.91 & 1.06 &   13 &   32 & &  110 &  148 &  194 &  122 &   33 &   76 &   11
\enddata
\tablenotetext{a}{No extinction correction applied.}
\tablenotetext{b}{Foreground screen correction applied to line emission.}
\tablenotetext{c}{Mixed gas/dust correction applied to line emission.}
\tablenotetext{d}{Percentage increase between first and second columns.}
\tablenotetext{e}{Percentage increase between first and third columns.}
\tablenotetext{f}{Foreground screen correction applied to both line emission and continuum.}
\tablenotetext{g}{Percentage increase between first and fourth columns.}
\end{deluxetable}

%% file: mackenty.tab5.tex
\begin{deluxetable}{lcccccccccc}
\tablecaption{Radio aperture data.\label{radioap}}
\tabletypesize{\scriptsize}
\tablewidth{0pt}
\tablehead{\colhead{name} & \colhead{$\alpha$ (J2000)} & \colhead{$\delta$ (J2000)} & \colhead{radius} & \colhead{log $F$\tablenotemark{a}} & \colhead{log $F$\tablenotemark{a}} & \colhead{log $F$\tablenotemark{b}} & \colhead{log $F$\tablenotemark{c}} & \colhead{$\alpha$\tablenotemark{d}} & \colhead{$\tau_{\rm rad}$\tablenotemark{e}} & \colhead{$\tau_{\rm Bal}$\tablenotemark{f}} \\
 & \colhead{12$^{\rm h}$15$^{\rm m}$+} & \colhead{36\arcdeg +} & \colhead{(arcsec)} & \colhead{(6 cm)} & \colhead{(20 cm)} & \colhead{(H$\alpha$)} & \colhead{(H$\alpha$)} &  &  & }
\startdata
   I-A & 39\fs 165 & 19\arcmin 33\farcs 97 &   5.00 &  -2.64 &  -2.48 & -11.91 & -11.73 &  -0.31 & 1.10 - 1.39 &        0.41 \\
   I-F & 39\fs 926 & 19\arcmin 34\farcs 03 &   4.00 &  -3.22 &  -2.51 & -12.12 & -11.99 &  -1.36 &         --- &        0.31 \\
   I-B & 40\fs 338 & 19\arcmin 29\farcs 11 &   5.00 &  -2.85 &  -2.43 & -12.05 & -11.91 &  -0.81 &         --- &        0.30 \\
 II-AE & 40\fs 937 & 19\arcmin 03\farcs 70 &   4.00 &  -2.70 &  -2.68 & -12.12 & -11.89 &  -0.04 &        0.86 &        0.53 \\
II-BCD & 40\fs 582 & 19\arcmin 10\farcs 74 &   4.00 &  -2.84 &  -2.62 & -12.08 & -11.96 &  -0.42 & 1.19 - 1.52 &        0.27 \\
\enddata
\tablenotetext{a}{log(radio flux) in Jy.}
\tablenotetext{b}{Measured log(\ha\ flux) in erg s$^{-1}$ cm$^{-2}$.}
\tablenotetext{c}{Extinction-corrected log(\ha\ flux) in erg s$^{-1}$ cm$^{-2}$ using the Balmer ratio and a foreground screen model.}
\tablenotetext{d}{Radio spectral index.}
\tablenotetext{e}{Optical depth at \ha\ measured from the radio / \ha\ ratio.}
\tablenotetext{f}{Optical depth at \ha\ measured from the Balmer ratio.}
\end{deluxetable}

%% file: mackenty.tab6.tex
\begin{deluxetable}{lcccc}
\tablecaption{Excitation ratios at the SNR location.\label{snrratios}}
\tabletypesize{\scriptsize}
\tablewidth{0pt}
\tablehead{\colhead{ratio} & \colhead{narrow comp.} & \colhead{wide comp.}}
\startdata
       \mbox{[N II]} $\lambda$6584 / \ha & $0.13\pm 0.01$ & $0.24\pm 0.05$ \\
  \mbox{[S II]} $\lambda$6717+6731 / \ha & $0.31\pm 0.01$ & $0.66\pm 0.09$ \\
      \mbox{[O III]} $\lambda$5007 / \ha & $1.71\pm 0.08$ & $1.55\pm 0.35$ \\
\enddata
\end{deluxetable}

%% file: mackenty.tab7.tex
\begin{deluxetable}{lrrrr}
\tablecaption{Integrated quantities as a function of pixel flux.\label{binned}}
\tabletypesize{\scriptsize}
\tablewidth{0pt}
\tablehead{\colhead{log} & \colhead{log} & \colhead{\% \ha\tablenotemark{c}} & \colhead{\% \mbox{[O III]}} & \colhead{exc.} \\
\colhead{(pixel flux)\tablenotemark{a}} & \colhead{(\%~$N$)\tablenotemark{b}} &  & \colhead{$\lambda$5007\tablenotemark{c}} & \colhead{ratio\tablenotemark{d}}}
\startdata
-17.0 or less  &   1.96 & 26.3 & 20.2 & 0.62 \\
-17.0 to -16.5 &   0.77 & 16.1 & 16.8 & 0.83 \\
-16.5 to -16.0 &   0.43 & 21.8 & 21.4 & 0.79 \\
-16.0 to -15.5 &  -0.02 & 22.1 & 23.1 & 0.84 \\
-15.5 to -15.0 &  -0.79 & 10.2 & 13.3 & 1.04 \\
-15.0 to -14.5 &  -1.76 &  2.9 &  4.5 & 1.27 \\
-14.5 or more  &  -3.02 &  0.5 &  1.1 & 1.69 \\
\enddata
\tablenotetext{a}{The flux for a single pixel is measured as the mean of \ha\ and \oiiir\ after a 3$\times$3 pixel box smoothing filter is applied to low intensity areas. The units are erg s$^{-1}$ cm$^{-2}$.}
\tablenotetext{b}{Percentage of pixels in that flux bin.}
\tablenotetext{c}{Percentage of the total flux in that bin.}
\tablenotetext{d}{Excitation ratio \mbox{[O III]} $\lambda$5007 / \ha\ for that flux bin.}
\end{deluxetable}